\documentclass[pre,aps,twocolumn,floatfix]{revtex4}
\usepackage{graphicx}

\newcommand{\kk}{{\vec{k}}}
\newcommand{\rr}{{\vec{r}}}
\newcommand{\LiMg}{{Li$_{0.7}$Mg$_{0.3}$~}}

\begin{document}

\title{Viscoelastic model for the dynamic structure factors of binary systems.}
\author{N. Anento}
\affiliation{Departament de F\'{\i}sica Fonamental. Universitat de Barcelona. 08028 Barcelona. SPAIN.}
\author{L.E. Gonz\'alez}
\author{D.J. Gonz\'alez}
\affiliation{Departamento de F\'{\i}sica Te\'orica. Universidad de Valladolid. 47011 Valladolid. SPAIN.}
\author{Y. Chushak}
\affiliation{
Department of Chemistry, University of Michigan, Ann Arbor, MI 48109, USA
}
\author{A. Baumketner\footnote{Permanent address: Institute for Condensed Matter Physics, 1 Svientsitsky Str., Lviv 79011, Ukraine.}}

\affiliation{Department of Chemistry and Biochemistry, University of California, Santa Barbara, CA 93106, USA}
\date{\today}

\begin{abstract}
This paper presents the viscoelastic model for the Ashcroft-Langreth dynamic 
structure factors of liquid binary mixtures. We also provide expressions for 
the Bhatia-Thornton dynamic structure factors and, within these expressions, 
show how the model reproduces both the dynamic and the self-dynamic structure 
factors corresponding to a one-component system in the appropriate limits 
(pseudobinary system or zero concentration of one component). 
In particular we analyze the behavior of the concentration-concentration 
dynamic structure factor and longitudinal current, and their corresponding 
counterparts in the one-component limit, namely, the 
self dynamic structure factor and self longitudinal current. 
The results for several lithium alloys with different ordering tendencies 
are compared with computer simulations data, leading to a good qualitative 
agreement, and showing the natural appearance in the model of the fast 
sound phenomenon.
\end{abstract}

\pacs{ {\bf PACS:} 61.20.Gy; 61.20.Lc; 61.25.Mv }

\maketitle

\section{Introduction.}
The development of inelastic neutron scattering (INS) techniques opened up, around 40 years ago, the experimental
study of the dynamic properties of several condensed matter systems, in particular of liquids. In principle
the total scattered intensity in an INS experiment includes both incoherent and coherent contributions, which are related respectively to the self-dynamic structure factors and the dynamic structure factors.
A clearcut separation of both contributions is not always possible and in the analysis of the raw data it is
useful to have simple models for the dynamic and/or self-dynamic structure factors  
in order to achieve such a separation through a numerical fitting procedure, and perform a proper interpretation of the experimental data. Even in those cases where there is coherent scattering only, it may happen that the particular behavior of the dynamic structure factor as a function of frequency obscures the analysis, for instance when no clear side peaks appear; in this case again the availability of models for fitting helps in the interpretation of the mechanisms controlling the behavior of the dynamic properties of the system. 
Similar problems are encountered when the dynamic properties of liquid systems are studied by either inelastic X-ray scattering (IXS) or molecular dynamics (MD) simulations. 
Even though IXS produces coherent scattering only, and MD provides a very detailed information of the properties of interest, nevertheless the interpretation of the numerical data obtained is greatly aided if theoretical models are available. 

In this respect, and for pure systems (one-component systems), a prominent role has been played by the so called viscoelastic model, introduced by Lovesey, which basically describes the dynamic structure factor as a sum of three Lorentzian functions of frequency, one of them representing particle diffusion and the other two describing damped propagation of collective excitations. This model applies for intermediate wave vectors, $k$, in-between those corresponding to a hydrodynamic behavior (low $k$, where the hydrodynamic model is applicable) and those corresponding to an ideal gas behavior (large $k$, where the free-particle model is correct).
A similar expression is also available for the self-dynamic structure factor, but its use has been much more scarce in the literature, although as we shall show below, the viscoelastic model for the self-dynamic structure factor in fact has very interesting properties that other models lack.

The case of mixtures is more complicated. Although both the hydrodynamic and the free-particle models are readily extended to mixtures, there is no well-behaved model so far to describe the intermediate $k$ range. A previous attempt to extend the viscoelastic model to liquid mixtures \cite{Chushak}, failed because some errors in the derivation made it incorrect, and therefore inapplicable. In particular the model did not recover the one-component case for pseudobinary systems, i.e. those systems which are in fact one-component, but with some particles labeled different from others (some of them are named type 1 and the rest type 2).

In this paper we extend the viscoelastic model to mixtures, and in particular to binary mixtures, in a consistent way which reduces to the correct one-component limit in the appropriate cases (pseudobinary mixture or zero concentration of one component). Moreover we give expressions for the so called Bhatia-Thornton dynamic structure factors, which are very useful when discussing ordering properties of binary systems. From these expressions it is easy to show the reduction to the one-component case, leading not only to the dynamic structure factor, but also to the self-dynamic structure factor of pure systems.

We compare the results of the model with MD data for three different type of systems: the first is a pseudobinary alloy, namely, pure liquid Li, the second corresponds to Li-Mg, which is a typical quasi ideal system, and finally the third one corresponds to Li$_4$Pb, which is an archetypical case for a class of compound forming alloys. The case of systems with tendency to phase separation has already been considered before \cite{NapoLiNa}, in the study of liquid Li$_{0.61}$Na$_{0.39}$, which is again a typical phase separating mixture. 
Although apparently trivial, the study of the pseudobinary case leads to interesting conclusions regarding the behavior of the self-dynamic structure factor, which vindicate the use of the viscoelastic model for the self-dynamic structure factor of one-component systems. 

\section{Formalism.}

\subsection{One-component system.}
Here we merely recall the expressions for the dynamic properties we are interested in for one-component systems.
The basic magnitude to be considered is the intermediate scattering function (ISF), $F(k,t)$, which 
describes the collective dynamic behavior of the system and is defined as
$F(k,t)=\langle \rho_{\kk}(t)\rho_{-\kk}(0) \rangle$, where $\rho_{\kk}(t)$ is the microscopic number density,
$$\rho_{\kk}(t) = \frac{1}{\sqrt N} \sum_{\ell=1}^{N} \exp \left[ i \; {\kk} \cdot {\rr}_{\ell}(t) \right] $$
of the system composed of $N$ particles at positions ${\rr}_{\ell}(t)$, which are enclosed in a volume $V$, so that the ionic number density
is $\rho=N/V$.
The initial value of the ISF is the static structure factor, $S(k)$, which is directly related to the pair distribution function, $g(r)$:
\begin{equation}
F(k,0)=S(k)=1+\rho \int d\rr \, \left(g(r)-1\right) \exp[-i\;\kk\cdot\rr] \,\,\, .
\end{equation}

A similar separation into two terms is also possible for all times,
\begin{equation}
F(k,t)=F_s(k,t)+F_d(k,t)
\end{equation}
where we have introduced the self intermediate scattering function (SISF), $F_s(k,t)$, and the distinct intermediate scattering function, $F_d(k,t)$, which obviously have the initial values $F_s(k,0)=1, F_d(k,0)=S(k)-1$.
The SISF is of interest by itself, since it is the basic magnitude that describes the one-particle 
dynamic behavior of the system.

The dynamic structure factor (DSF), $S(k,\omega)$, and the self dynamic structure factor (SDSF), $S_s(k,\omega)$, are obtained from the corresponding intermediate scattering functions by passing to the Fourier domain:

\begin{equation}
\label{dsf_1comp} 
S(\kk, \omega) = \frac{1}{2 \pi} \int_{-\infty}^{\infty} 
dt \; e^{i\;\omega \; t} \; F(\kk, t) = 
\frac{1}{\pi} \; {\it Re} \;  \tilde F(\kk , z= -i \omega) 
\end{equation}

\noindent where {\it Re} stands for the real part and $\tilde F(\kk, z)$ 
is the Laplace transform of $F(\kk, t)$, i.e.  

\begin{equation}
\label{laplace_1comp}
\tilde F(\kk, z) = \int_0^{\infty} dt \; e^{-zt} \; F(\kk, t)
\end{equation}

The memory functions of the ISF, $M(k,t)$, and of the SIFS, $M_s(k,t)$, are then introduced through the generalized Langevin equations, which read in the time domain and in the Laplace domain:
\begin{eqnarray}
\label{Langevin_1comp}
\frac{d}{dt}F(\kk, t) = - \; \int_0^t 
d \tau \; M(\kk, \tau) \; F(\kk, t-\tau)
\\
\label{Fkz_1comp}
\tilde{F}(\kk, z) = \left[ z  
+ \tilde {M}(\kk, z) \right]^{-1} \; F(\kk,t=0)
\end{eqnarray}  
with equivalent equations for the self counterparts. The higher order memory functions are introduced exactly in the same way: the second order memory functions $N(k,t)$ and $N_s(k,t)$ are the memory functions of $M(k,t)$ and $M_s(k,t)$ respectively, while the third order memory functions, $K(k,t)$ and $K_s(k,t)$ are the memory functions of $N(k,t)$ and $N_s(k,t)$ respectively.
The initial values of the memory functions which appear in the Laplace formulation of the generalized Langevin equation are easily determined in terms of the $n$th frequency moments of the DSF and SDFS, $\omega^n(\kk)=\int_{-\infty}^{\infty} \omega^n \; S(\kk, \omega) \; d\omega$,

\begin{eqnarray}
\label{Mktz_1comp}
M(\kk, t=0) & = & \omega^2 (\kk) \; 
 \omega^0 (\kk)^{-1}
\\
\label{Nktz_1comp}
N(\kk, t=0) & = & \omega^4 (\kk) \; 
\omega^2 (\kk)^{-1} \; - \; 
\omega^2 (\kk) \; 
\omega^0 (\kk)^{-1}
\end{eqnarray}
with again equivalent expressions for the self counterparts. Moreover these frequency moments up to the fourth
can be determined from the knowledge of only the temperature $T$, the atomic mass, the interatomic pair potential, $\phi(r)$, and the static structure, i.e., $g(r)$ and $S(k)$.  

A useful feature of a memory function is that it decays in time more rapidly than the function from which it originates. Based on this, it seems plausible that at some level of the hierarchy of memory functions an approximation where the memory function is just a Dirac delta function at $t=0$ should be a good ansatz.
In the viscoelastic model the level at which this ansatz is taken is the third, i.e., it is assumed that
\begin{eqnarray}
K(\kk,t)=K(\kk) \delta(t) \quad & \Rightarrow & \quad \tilde{K}(\kk,z)=K(\kk) \nonumber \\
K_s(\kk,t)=K_s(\kk) \delta(t) \quad & \Rightarrow & \quad \tilde{K}_s(\kk,z)=K_s(\kk)
\end{eqnarray}

Explicit expressions of $K(\kk)$ and $K_s(\kk)$ in terms of the same magnitudes as the frequency moments can
be obtained \cite{Lovesey,Copley&Lovesey} by imposing that in the free particle limit ($k \to \infty$) the theory recovers the known exact $S(k,\omega=0)=S_s(k,\omega=0)$. Further details are given in the next section.

Introducing in eq.~(\ref{Fkz_1comp}) the higher order memory functions up to the third it is found that
\begin{equation}
\tilde{F}(\kk,z)=\frac{F(\kk,t=0)}{ z + \displaystyle\frac{M(\kk,t=0)}{ z+ \displaystyle\frac{N(\kk,t=0)}{z+K(\kk)}}} = \frac{P_2(\kk,z)}{P_3(\kk,z)}
\end{equation}
where $P_n(\kk,z)$ denotes a real polynomial in $z$ of degree $n$. Making a partial fraction decomposition of this expresion, and denoting by $z_i$ the roots of $P_3(\kk,z)$, which, by the way, are the same as the roots of
$z+\tilde{M}(\kk,z)$, we have
\begin{equation}
\tilde{F}(\kk,z)=\sum_{i=1}^3\frac{A_i}{z-z_i} \quad \Rightarrow \quad F(\kk,t)=\sum_{i=1}^3 A_i \exp[z_i t]
\end{equation}
where the $z_i$ and the $A_i$ appear either as real or in complex conjugate pairs. Therefore either
the three roots are real or one is real and the other two are a complex conjugate pair. In all practical situations the latter is the case, and then the roots are interpreted as describing a diffusive mode
and a pair of damped propagating modes, much the same as in the hydrodynamic model, although for $k$ values 
outside this regime. The DSF is then a sum of three Lorentzian functions, one centered at
$\omega=0$ which corresponds to the real root, and the other two centered at the imaginary parts of
the complex conjugate pair.

At small $k$, the functional form of the dynamic magnitudes 
within  the viscoelastic model coincides with that of the hydrodynamic 
model, which is known to be accurate in this region. 
Futhermore, the viscoelastic modes behave (as a function of $k$) much in 
same way as the hydrodynamic modes. However, in spite of the previous 
similarities, there are basic differences between the viscoelastic and the 
hydrodynamic models. 
They are better understood if  the derivation of both models 
is made by an alternative route using the generalized modes approach 
\cite{Schepper}. Here one considers the equations of motion of several dynamic
magnitudes, namely temperature (or energy) fluctuations, density fluctuations,
current fluctuations and stress tensor fluctuations. The viscoelastic model 
follows from considering couplings among the last three variables and ignoring
their coupling with temperature fluctuations. The hydrodynamic model follows
from considering couplings between the first three magnitudes   
and ignoring their coupling to the stress tensor fluctuations.
A parameter quantifying the coupling between density and temperature 
fluctuations is the specific heat ratio $\gamma=C_p/C_v$. If $\gamma\sim 1$ the 
coupling is weak and the viscoelastic model is expected to be accurate; 
otherwise the model is expected to fail especially at small $k$.

Within the viscoelastic model, the analytical 
structure of the self functions, i.e. 
$F_s(k,z)$, the SISF and the SDSF, 
is the same as that of the equivalent collective 
counterparts. However, the amplitude of the real 
coefficient associated to the real root is usually much greater 
than the amplitudes of the complex terms  
and therefore the associated propagating 
modes in the SDSF have often been neglected when the viscoelastic model 
has been 
applied to study the dynamic properties of one-component liquids. 
It may be argued whether these modes are real or a mere artifact of the model 
because the SDSF describe the one-particle behavior of the system. 
However, the close connection 
between collective and single particle properties in dense systems 
could induce  the appearance of these modes.
In any case, we stress  that, to our knowledge, no detailed study of this 
possibility has yet been carried out.

\subsection{Binary system: Ashcroft-Langreth partials.}

\label{theory}

The generalization of the foregoing formalism to binary systems is  straightforward. We consider $N$ particles in a volume $V$, therefore with number density $\rho=N/V$, composed of two species ($i=1,2$) which are characterized by the number of particles $N_i$, or concentration $x_i=N_i/N$, the atomic masses $m_i$, and
ionic partial number densities $\rho_i=x_i \rho$. The interaction between particles of type $i$ and $j$ is described by effective interatomic pair potentials $\phi_{ij}(r)$ and the static structure by the partial 
pair distribution functions $g_{ij}(r)$.

We start from the Fourier transform of the partial microscopic number 
densities, defined as  

\begin{equation}
\rho_{\kk}^{(j)}(t) = \frac{1}{\sqrt N_j} \sum_{\ell=1}^{N_j} 
\exp \left[ i \; {\kk} \cdot {\rr}_{\ell}^{(j)}(t) \right] \qquad \qquad (j=1,2) 
\end{equation} 

\noindent from which the partial 
intermediate scattering functions, $F_{ij}(\kk, t)$ 
are obtained as 

\begin{equation}
F_{ij}(\kk, t) = < \rho_{\kk}^{(i)}(t)\; \rho_{-\kk}^{(j)}(0) > = 
\left[ {\bf F}(\kk, t) \right]_{ij}
\end{equation}

\noindent where the last equation defines the 2 x 2 matrix ${\bf F}(\kk, t)$. The initial value of this 
matrix reduces to the matrix of Ashcroft-Langreth (AL) 
partial structure factors,
\begin{equation}
{\bf F}(\kk,t=0)={\bf S}(\kk)
\end{equation}
where
\begin{eqnarray}
[{\bf S}(\kk)]_{ij}& =& S_{ij}(\kk) \nonumber \\
& = & \delta_{ij}+(x_i x_j)^{1/2} \rho \int d\rr \, \left(g_{ij}(r)-1\right) \exp[-i\;\kk\cdot\rr]
\nonumber \\
& & 
\end{eqnarray}

As in the case of one-component systems, a similar separation into two terms is possible for all times, defining
the self and distinct parts of the partial ISFs:
\begin{eqnarray}
{\bf F}(\kk,t)&=&{\bf F_s}(\kk,t) + {\bf F_d}(\kk,t) \nonumber \\
F_{ij}(\kk,t)&=&\delta_{ij}F_{s,j}(\kk,t)+(x_i x_j)^{1/2} F_{d,ij}(\kk,t)
\label{def_isfs}
\end{eqnarray}

Equations (\ref{dsf_1comp})-(\ref{Nktz_1comp}) remain formally the same, although now all the magnitudes
in the equations are matrices, the products are to be understood as matrix products and the exponent -1
means matrix inverse. For instance, the memory function matrix obeys in the time and the Laplace domain
the equations
\begin{equation}
\label{Langevin}
\frac{d}{dt}{\bf F}(\kk, t) = - \; \int_0^t 
d \tau \; {\bf M}(\kk, \tau) \; {\bf F}(\kk, t-\tau)
\end{equation}

\begin{equation}
\label{Fkz}
\tilde{{\bf F}}(\kk, z) = \left[ z {\bf I}  
+ \tilde{{\bf M}}(\kk, z) \right]^{-1} \; {\bf F}(\kk,t=0)
\end{equation}  
where ${\bf I}$ is the unit matrix.

The matrices of frequency moments are given by \cite{Price&Copley}
\begin{equation}
\label{Omega02}
\left[ {\bf \omega}^0 (\kk) \right]_{ij} = S_{ij}(\kk) \qquad
\left[ {\bf \omega}^2 (\kk) \right]_{ij} = \delta_{ij} \; k^2 \frac{k_B T}{m_i}
\end{equation}

\begin{eqnarray}
\label{Omega4}
& & \left[ {\bf \omega}^4 (\kk) \right]_{ij}  = \nonumber \\
& &\delta_{ij} \; 
k^2 \frac{k_B T}{m_i m_j} \; \left[ 3 k^2 k_B T + \sum_{\ell} x_{\ell} \rho 
\int d \rr \; g_{i\ell}(\rr) \; \frac{\partial^2 
\varphi_{i\ell}(\rr)}{\partial z^2} \; \right] \nonumber \\
& & - (x_i x_j)^{1/2} k^2 \frac{k_B T}{m_i m_j} \;  \rho 
\int d \rr  \; g_{ij}(\rr) \cos(\kk \rr) \; 
\frac{\partial^2 \varphi_{ij}(\rr)}{\partial z^2}  \nonumber \\
& & = k^2 \frac{k_BT}{m_im_j} \left\{ \delta_{ij} \left[ 3k^2 k_BT + \sum_{\ell} x_{\ell}
\chi_{i\ell}(k)\right] \right. \nonumber \\
& & - \left. (x_ix_j)^{1/2}\chi_{L;ij}(k) \right\}
\end{eqnarray}

\noindent where $T$ is the temperature, $k_B$ is Boltzmann's constant, $z$ denotes the direction of $\kk$ and
the last equality defines the functions $\chi_{ij}$ and $\chi_{L;ij}$.

The hierarchy of memory function matrices is again truncated at the third level, setting
\begin{equation}
{\bf K}(\kk,t)={\bf K}(\kk) \delta(t)  \qquad \Rightarrow \qquad \tilde{{\bf K}}(\kk,z)={\bf K}(\kk) 
\end{equation}
so that explicitly we have for the ISF matrix and the first and second order memory function matrices
the relation (\ref{Fkz}) and
\begin{equation}
\label{Mkz}
\tilde{{\bf M}}(\kk, z) = \left[ z {\bf I}  
+ \tilde{{\bf N}}(\kk, z) \right]^{-1} \; {\bf M}(\kk,t=0)
\end{equation}  
\begin{equation}
\label{Nkz}
\tilde{{\bf N}}(\kk, z) = \left[ z {\bf I}  
+ {\bf K}(\kk) \right]^{-1} \; {\bf N}(\kk,t=0)
\end{equation}  
with the initial values of the memory function matrices given in terms of the frequency moments matrices
as in eqs. (\ref{Mktz_1comp}) and (\ref{Nktz_1comp}).
In order to provide explicit expressions for the matrix elements of ${\bf K}(\kk)$ we follow Lovesey's arguments \cite{Lovesey,Copley&Lovesey}. Setting $z=0$ in eqs. (\ref{Mkz}) and (\ref{Nkz}) we have
\begin{eqnarray}
\tilde{{\bf M}}(\kk, z=0) &=& \tilde{{\bf N}}(\kk, z=0)^{-1} \; {\bf M}(\kk,t=0) \nonumber \\
& = &
\left[ {\bf K}(\kk) ^{-1} \; {\bf N}(\kk,t=0) \right]^{-1}  \; {\bf M}(\kk,t=0) \nonumber \\ 
& = &
{\bf N}(\kk,t=0)^{-1} \; {\bf K}(\kk) \; {\bf M}(\kk,t=0)
\end{eqnarray}
and therefore
\begin{equation}
\label{Krelax}
{\bf K}(\kk) = {\bf N}(\kk,t=0) \; \tilde{{\bf M}}(\kk, z=0) \; {\bf M}(\kk,t=0)^{-1}
\end{equation}

The value of $\tilde{{\bf M}}(\kk, z=0)=\int_0^{\infty} dt \; {\bf M}(\kk,t)$ is then assumed to be rather insensitive to the detailed shape of the $M_{ij}(\kk,t)$ and therefore, a reasonable estimate can be obtained from an approximation that fulfills its correct short time behavior obtained by a simple Taylor expansion, namely,
\begin{equation} 
\label{Mshortt}
{\bf M}(\kk, t)  =  \left[ {\bf I} - 
\frac{t^2}{2} {\bf N}(\kk, t=0)+ ......\right] {\bf M}(\kk, t=0) \,\,\, .
\end{equation}
The specific approximation we make for ${\bf M}(\kk,t)$ to perform the integral is 
\begin{equation}
\label{Mapprox}
{\bf M}(\kk, t) \approx \exp \left[ -\frac{t^2}{2} {\bf N}(\kk, t=0)\right] 
\; {\bf M}(\kk, t=0) \,\,\, . 
\end{equation}
It is precisely at this point where the previous attempt to generalize the
viscoelastic model \cite{Chushak}
failed, since a similar approximation was made at the level of the
matrix elements $M_{ij}(\kk,t)$ only,  and not for the matrix 
${\bf M}(\kk,t)$ itself, which is obviously wrong because we are 
dealing with matrix products.

Within this approximation $\tilde{{\bf M}}(\kk,z=0)$ would be given by the time integral of the 
exponential function times the matrix ${\bf M}(\kk, t=0)$. Trying to correct the inaccuracies that
might have been introduced by this approximation, this value is premultiplied by a matrix ${\bf \Xi}$
of constants to be determined later. In this way an explicit expression for
${\bf K}(\kk)$ is found in terms of the initial values of the second order memory functions:
\begin{equation}
{\bf K}(\kk) = {\bf N}(\kk,t=0) \; {\bf \Xi} \; \int_0^{\infty} dt\,
\exp \left[ -\frac{t^2}{2} {\bf N}(\kk, t=0)\right] \,\,\, .
\end{equation}
The determination of ${\bf \Xi}$ is carried out by imposing that in the free particle limit ($k\to\infty$) the dynamic structure factor matrix evaluated at zero frequency recovers the exact free gas result, i.e.,
\begin{equation}
[{\bf S}(\kk,\omega=0)]_{ij}=\delta_{ij} \; \left( \frac{1}{2 \pi k^2} 
\frac{m_i}{k_B T} \right)^{1/2}
\end{equation}
leading to the result ${\bf \Xi} = (2 \sqrt{2}/\pi) {\bf I}$. 
In the Appendix we give details of this derivation and explicit expressions for the matrix elements of 
${\bf K}(\kk)$.

Turning now to the analytic behavior of $\tilde{{\bf F}}(\kk,z)$ we see in eq.~(\ref{Fkz}) that it is determined by the inverse of the matrix $[z{\bf I} + \tilde{{\bf M}}(\kk,z)]$, namely the transpose of the adjoint divided by the determinant. Therefore, 
all the $\tilde{F}_{ij}(k,z)$ have the same analytic behavior, which  
is determined by the solutions of the equation
\begin{equation}
\det \left[ z {\bf I} + \tilde {\bf M}(\kk, z) \right] = 0
\end{equation} 
As in the one-component case, this determinant is a real rational function, whose form is obtained
by writing down explicitly the equations for the memory function matrices up to the third, leading now to an expression of the type $P_6(\kk,z)/P_4(\kk,z)$. 
Consequently there will be six roots, which in principle may appear
(together with the corresponding amplitude matrices) as 6 real values, 4 real values and a complex conjugate pair, 2 real values and 2 complex conjugate pairs, or 3 complex conjugate pairs. Therefore we have
\begin{equation}
\tilde{{\bf F}}(\kk,z)=\sum_{i=1}^6\frac{1}{z-z_i}{\bf A_i} \qquad  
{\bf F}(\kk,t)= \sum_{i=1}^6 {\bf A_i} \exp[z_i t] 
\end{equation}
where the ${\bf A_i}$ are the $\kk$-dependent 2 x 2 amplitude matrices.
In the actual calculations made for liquid Li-Na \cite{NapoLiNa}, 
as well as those carried out in this paper, we found for all $k$ two real roots and 
two pairs of complex conjugate roots, at variance with some other calculations
where one of the complex conjugate pair transforms into two real roots below
a certain small $k$ value \cite{Schepper88,Bryk97}.
All roots have negative real parts \cite{Schepper88}, so we can 
rewrite the previous equation as
\begin{eqnarray}
\tilde{F}_{ij}(\kk,z)&=&\frac{A_{ij}}{z-z^{(1)}}+\frac{A_{ij}^*}{z-z^{(1)*}}+
\frac{B_{ij}}{z-z^{(2)}}+\frac{B_{ij}^*}{z-z^{(2)*}} \nonumber \\
& + &\frac{C_{ij}}{z+z^{(3)}}+
\frac{D_{ij}}{z+z^{(4))}} \,\,\,\, , 
\end{eqnarray}
where the $\kk$-dependent coefficients $A_{ij}$ and $B_{ij}$ are complex, while $C_{ij}$ and $D_{ij}$ are real.
The complex roots,  
$z^{(j)}(k)= - \Gamma^{(j)}(k) \pm i \omega^{(j)}_s(k), \; (j=1,2)$ 
describe propagation, where the real part  $\Gamma^{(j)}(k)$, represents 
damping whereas the imaginary part $\omega^{(j)}_s(k)$,
represents the propagation. A set of two complex 
conjugate roots represents propagation in opposite directions. The real 
roots, $-z^{(j)}(k), \; (j=3,4)$ stand for purely diffusive processes.  
Out of the six roots, only four go to zero when $k \to 0$; this which coincides 
with the behavior of the four hydrodynamic roots 
 predicted by the hydrodynamic equations for binary mixtures.
Again we stress that although the small $k$ behavior of the 
four viscoelastic modes is similar to  
the hydrodynamic ones, there are quantitative differences between them 
based on the same reason
as in the one-component case, namely the neglect of coupling with the energy
fluctuations. As in one-component liquids, 
the viscoelastic model is expected to work for systems 
where this coupling is weak, i.e., those with 
specific heat ratio $\gamma\sim 1$.
The other two viscoelastic roots have a finite value 
when $k \to 0$, while the corresponding amplitudes vanish in
this limit; this is the typical behavior of kinetic modes. 
The six roots lead to a  $S_{ij}(k, \omega)$ given as a  
sum of six Lorentzians, two centered at $\omega = 0$ and the other four at 
$\omega = \pm \omega_s^{(j)}(k) , \; (j=1,2)$.

\subsection{Binary system: Bhatia-Thornton partials.}

An alternative description of the structure of binary alloys is provided by 
the so called Bhatia-Thornton (BT)
functions. The BT static partial structure factors, namely, number-number
$S_{NN}(k)$, number-concentration $S_{Nc}(k)$, and concentration-concentration $S_{cc}(k)$ partial structure factors describe the correlations among fluctuations in number density (irrespective of chemical composition)
and concentration density, and are linear combinations of the AL partial structure factors $S_{ij}(k)$. These relations are most compactly written if one defines the matrix
${\cal S}(k)$ of modified BT partial structure factors in terms of the matrix ${\bf S}(k)$ of AL partial structure factors:
\begin{equation}
{\cal S}  =  \bf X \; {\bf S} \; {\bf X} 
\end{equation}
where
\begin{equation}
{\cal S}(k)=\left( \begin{array}{cc} S_{NN}(k) & S_{Nc}(k)/(x_1x_2)^{1/2} \\
                 S_{cN}(k)/(x_1x_2)^{1/2} & S_{cc}(k)/(x_1x_2) \end{array} \right) 
\end{equation}
and 
\begin{displaymath}
{\bf X}={\bf X}^{-1}=\left( \begin{array}{cc} \sqrt{x_1} & \sqrt{x_2} \\
                                              \sqrt{x_2} & -\sqrt{x_1} \end{array} \right)
\end{displaymath}

All other BT magnitudes (intermediate scattering functions, first, second and third memory functions, frequency moments, etc.) are defined in the same way, pre- and post- multiplying the 
corresponding matrix of AL magnitudes by the matrix ${\bf X}$; for instance,
the BT partial ISFs are defined by the matrix ${\cal F}(\kk,t)$,
\begin{equation}
{\cal F}(\kk,t) = {\bf X} \;  {\bf F}(\kk,t) \; {\bf X} = {\cal F}_s(\kk,t) + {\cal F}_d(\kk,t)
\end{equation}
which leads to (dropping the arguments of the functions)
\begin{eqnarray}
F_{NN}&=&\left\{ x_1 F_{s,1} + x_2 F_{s,2} \right\} \nonumber \\
& + & \left\{ x_1^2 F_{d,11} + 2 x_1x_2 F_{d,12} + x_2^2 F_{d,22} \right\} 
\end{eqnarray}
\begin{eqnarray}
&&\displaystyle\frac{F_{Nc}}{(x_1x_2)^{1/2}}  =  \left\{ (x_1x_2)^{1/2} (F_{s,1}-F_{s,2}) \right\} \nonumber \\
&&+ \left\{ (x_1x_2)^{1/2} \left[ x_1(F_{d,11}-F_{d,12})+x_2(F_{d,12}-F_{d,22})\right] \right\} \nonumber \\
&& 
\end{eqnarray}
and
\begin{eqnarray}
\displaystyle\frac{F_{cc}}{x_1x_2} & = & \left\{ x_2 F_{s,1}+x_1 F_{s,2}) \right\} \nonumber \\
&+& \left\{ x_1x_2 \left[F_{d,11}-2F_{d,12}+F_{d,22})\right]\right\}
\end{eqnarray}
where the first braces in each equation enclose the BT SISFs 
and the second braces enclose the BT distinct ISFs.

The viscoelastic model for the BT functions is then defined by the relations
\begin{eqnarray}
\tilde{\cal F}(\kk,z)=\left[z{\bf I}+\tilde{\cal M}(\kk,z)\right]^{-1} {\cal F}(\kk,t=0) \\
\tilde{\cal M}(\kk,z)=\left[z{\bf I}+\tilde{\cal N}(\kk,z)\right]^{-1} {\cal M}(\kk,t=0) \\
\tilde{\cal N}(\kk,z)=\left[z{\bf I}+{\cal K}(\kk)\right]^{-1} {\cal N}(\kk,t=0) \\
{\cal K}(\kk)={\cal N}(\kk,t=0) \; {\bf \Xi} \int dt \, \exp\left[-\frac{t^2}{2} {\cal N}(\kk,t=0)\right] \\
{\cal F}(\kk,t=0)=\Omega^0(\kk)={\cal S}(k) \\
{\cal M}(\kk,t=0)=\Omega^2(\kk) \Omega^0(\kk)^{-1} \\
{\cal N}(\kk,t=0)=\Omega^4(\kk) \Omega^2(\kk)^{-1} - \Omega^2(\kk) \Omega^0(\kk)^{-1} \\
\Omega^n(\kk)={\bf X} \; \omega^n(\kk) \; {\bf X}
\end{eqnarray}
explicitly
\begin{eqnarray}
\Omega^2_{NN}=k^2k_BT\left(\frac{x_1}{m_1}+\frac{x_2}{m_2}\right) \nonumber \\
\Omega^2_{cc}=k^2k_BT\left(\frac{x_1}{m_2}+\frac{x_2}{m_1}\right) \nonumber \\
\Omega^2_{Nc}=k^2k_BT(x_1x_2)^{1/2}\left(\frac{1}{m_1}-\frac{1}{m_2}\right)
\end{eqnarray}
and
\begin{widetext}
\begin{eqnarray}
\frac{\Omega^4_{NN}}{k^2k_BT}&=& \left(\frac{x_1}{m_1^2}+\frac{x_2}{m_2^2}\right)3k^2k_BT
+ x_1x_2\left(\frac{1}{m_2}-\frac{1}{m_1}\right)^2 \chi_{12} \nonumber \\
& + &
 \left( \frac{x_1^2}{m_1^2} \left[\chi_{11}-\chi_{L;11}\right]
+ 2 \frac{x_1x_2}{m_1m_2} \left[\chi_{12}-\chi_{L;12}\right] 
+ \frac{x_2^2}{m_2^2} \left[\chi_{22}-\chi_{L;22}\right] \right)
\nonumber \\
\frac{\Omega^4_{cc}}{k^2k_BT}&=& \left(\frac{x_1}{m_2^2}+\frac{x_2}{m_1^2}\right)3k^2k_BT
+ \left(\frac{x_1}{m_2}+\frac{x_2}{m_1}\right)^2 \chi_{12} \nonumber \\
& + &
 x_1x_2 \left( \frac{1}{m_1^2} \left[\chi_{11}-\chi_{L;11}\right]
- \frac{2}{m_1m_2} \left[\chi_{12}-\chi_{L;12}\right] 
+ \frac{1}{m_2^2} \left[\chi_{22}-\chi_{L;22}\right] \right)
\nonumber \\
\frac{\Omega^4_{Nc}}{k^2k_BT(x_1x_2)^{1/2}}&=& \left(\frac{1}{m_1^2}-\frac{1}{m_2^2}\right) 3k^2k_BT
+x_1\left( \frac{\chi_{11}}{m_1^2}-\frac{\chi_{12}}{m_2^2}\right) 
+x_2\left(\frac{\chi_{12}}{m_1^2}-\frac{\chi_{22}}{m_2^2}\right) \nonumber \\
& - & x_1\left(\frac{\chi_{L;11}}{m_1^2}-\frac{\chi_{L;12}}{m_1m_2}\right)
- x_2\left(\frac{\chi_{L;12}}{m_1m_2}-\frac{\chi_{L;22}}{m_2^2}\right)
\end{eqnarray}
\end{widetext}

\subsection{Reduction to the one-component case.}

In the case of a pseudobinary mixture, when all the particles are of the same type but are labelled different,
all the partial pair distribution functions reduce to that of the real one-component liquid, i.e., $g_{11}(r)=g_{22}(r)=g_{12}(r)=g(r)$.
However, the AL partial structure factors, do not coincide with that of 
the real one-component system, $S(k)$,  but 
\begin{displaymath}
S_{ij}(k)=\delta_{ij}+(x_ix_j)^{1/2} \left[ S(k)-1 \right] \,\,\, .
\end{displaymath}
Explicitly we have 
\begin{eqnarray}
S_{11}(k)&=&x_2+x_1S(k) \nonumber \\
S_{22}(k)&=&x_1+x_2S(k) \nonumber \\
S_{12}(k)&=&-(x_1x_2)^{1/2}+(x_1x_2)^{1/2}S(k) \,\, .
\label{static_pseudobinary}
\end{eqnarray}
The BT partial structure factors, on the other hand, behave more simply, 
because we now 
have  $S_{NN}(k)=S(k)$, $S_{Nc}(k)=0$ and $S_{cc}(k)/(x_1x_2)=1$.

In many one-component systems (in particular liquid metals near the triple 
point) the value of $S(k)$ for small $k$ is rather small, of 
the order $2-3 \times 10^{-2}$, reflecting a small value of the 
isothermal compressibility.
For a pseudobinary system, we therefore obtain  that, except for very 
dilute mixtures, the small $k$ values of the partial static structure factors 
are dominated by the first terms of eqns.~(\ref{static_pseudobinary}), which 
take on significantly higher values than that of the structure factor of 
the one-component liquid. 

The situation concerning the ISFs is similar. According to their definition, see eqn.~(\ref{def_isfs}), we have $F_{s,1}(\kk,t)=F_{s,2}(\kk,t)=F_s(\kk,t)$, which is the SISF of the real one-component liquid, and 
$F_{d,11}(\kk,t)=F_{d,22}(\kk,t)=F_{d,12}(\kk,t)=F_d(\kk,t)$,
which is the distinct ISF of the real one-component liquid.
Therefore the AL partial ISFs are given by
\begin{displaymath}
F_{ij}(\kk,t)=\delta_{ij}F_s(\kk,t)+(x_ix_j)^{1/2} F_d(\kk,t) \,\,\, ,
\end{displaymath}
and explicitly,
\begin{eqnarray}
F_{11}(\kk,t)&=&x_2F_s(\kk,t)+x_1F(\kk,t) \nonumber \\
F_{22}(\kk,t)&=&x_1F_s(\kk,t)+x_2F(\kk,t) \nonumber \\
F_{12}(\kk,t)&=&-(x_1x_2)^{1/2}F_s(\kk,t)+(x_1x_2)^{1/2}F(\kk,t) \,\, , \nonumber \\
& &
\end{eqnarray}
which is an ackward relation that induces a behavior of the partial ISFs very
different from that of the real one-component liquid. On the other 
hand the BT partial ISFs become
\begin{eqnarray}
F_{NN}(\kk,t)&=&(x_1+x_2)F_s(\kk,t) + (x_1+x_2)^2 F_d(\kk,t)\nonumber \\
& =& F_s(\kk,t)+F_d(\kk,t)=F(\kk,t) \\
F_{Nc}(\kk,t)&=& 0 \\
F_{cc}(\kk,t)/(x_1x_2)&=& (x_1+x_2)F_s(\kk,t)  = F_s(\kk,t) \,\,\, ,
\end{eqnarray}
that is, the cross function vanishes, the number-number function reduces to the ISF of the real one-component
liquid, and the concentration-concentration function, properly normalized, reduces to the SISF of the
real one-component system.

Coming to the viscoelastic model, in particular in the BT formulation, we note that all the magnitudes are
determined by the frequency moments matrices. In a pseudobinary system we have $m_1=m_2$, all the $\chi_{ij}$ reduce to
$\chi$ which is obtained by equalling 
$g_{ij}(r)=g(r)$ and $\phi_{ij}(r)=\phi(r)$; also by  
the same procedure, all the
$\chi_{L;ij}$ reduce to $\chi_L$. Then all the frequency moments 
matrices in the
BT formulation become diagonal, and therefore the same 
occurs to all the other matrices, so that the
matrix operations in fact become the same operations on the 
diagonal elements of the matrices, i.e., the
$NN$ and the $cc$ functions.

In particular we have 
\begin{eqnarray}
\Omega^4_{NN}=\frac{k^2k_BT}{m^2}\left[3k^2k_BT + \chi(k) -\chi_L(k) \right] \\
\Omega^4_{cc}=\frac{k^2k_BT}{m^2}\left[3k^2k_BT + \chi(k) \right]
\end{eqnarray}
which are respectively the 4th frequency moments of the DSF and the SDSF of the one-component system. Likewise
\begin{equation}
\Omega^2_{NN}=\frac{k^2k_BT}{m}=\Omega^2_{cc}
\end{equation}
which again are the second frequency moments of the DSF and the SDSF of the one-component system, and
as we stated before $\Omega^0_{NN}=S_{NN}(k)=S(k)$ and $\Omega^0_{cc}=S_{cc}(k)/(x_1x_2)=1$, which are the corresponding zeroth frequency moments of the DSF and the SDSF of the one-component system, respectively.
 
Therefore, the present formulation of the viscoelastic model for the collective dynamic properties 
of binary mixtures recovers, in the case of a pseudobinary system, not only the viscoelastic model
for the collective properties of the underlying one-component system, through the number-number BT 
functions, but also the viscoelastic model 
for its single-particle properties, through the concentration-concentration 
functions. Moreover, the determinant that leads to the different modes 
factorizes into two terms, one that includes the modes of the DSF with the 
other term accounting for  the modes of the SDSF of the one-component 
system; this implies that both sets of modes  are decoupled.

\section{Results.}

In this section we study three systems and compare the results of the 
viscoelastic model with those
obtained by molecular dynamics (MD) simulations.

The first is a pseudobinary system, in which all particles 
are in fact the same type, but half of them are labeled as 1 and the 
other half as 2 ($x_1=x_2=0.5$). 
Specifically the system is representative of $^7$Li at $T=725$ K 
and $\rho=0.042$ \AA$^{-3}$, and the effective pair potentials 
used (all the three identical) were
obtained using the empty core pseudopotential \cite{Ash_core}, with 
core radius $1.44$ \AA. Some MD results for this system have been 
reported elsewhere \cite{NapoJoanOlgaQuim} and are extended here. They 
have been obtained using 668 particles in a cubic 
box with periodic boundary conditions. 

The other two cases correspond to two Li-based alloys, Li$_{0.7}$Mg$_{0.3}$, which is a typical quasi-ideal mixture, and Li$_4$Pb, which is a reference case for a class of compound forming alloys with pre-ionic ordering. The temperatures and densities are $T=887$ K and $\rho=0.040711$ \AA$^{-3}$ for \LiMg and $T=1085$ K and $\rho=0.04558$ \AA$^{-3}$ for Li$_4$Pb.
The interatomic pair potentials were obtained in the case of \LiMg within the neutral pseudoatom method \cite{strucLiNaMg} while in the case of Li$_4$Pb were taken from Copestake {\em et al} \cite{Copestake}.
The collective propeties of the Li$_{0.7}$Mg$_{0.3}$ alloy have been 
studied by Anento {\em et al} \cite{Napo_PRB62,Napo_RM}, and those 
of Li$_4$Pb by several authors \cite{Bosse,Marta}, the latter being 
the first system where the fast sound phenomenon was observed, which 
consists in the appearance of a peak in the Li-Li dynamic structure 
factor, absent in the Pb-Pb one, with a very high frequency similar to that 
of pure Li. In this paper we include MD results obtained 
for both systems, in the case of \LiMg with 570 particles 
and in the case of Li$_4$Pb with 600 particles for most $k$ values 
and with 11000 particles for the smallest ones ($k_{\rm min}=0.10$ \AA$^{-1}$).

The difference between the AL ISFs of the two type of systems is
striking: while Li$_4$Pb has $F_{ij}(k,t)$, and in particular the Li-Li 
partial, that decay more or less rapidly with time, similar to the 
behavior of one-component systems, Li$_{0.7}$Mg$_{0.3}$ on the 
other hand shows $F_{ij}(k,t)$ which have much larger magnitudes and 
decay much more slowly, in practice making very difficult the 
Fourier Transformation that leads to the DSFs.

A proper description of such different behaviors, along with the case already studied of phase-separating systems like Li$_{0.69}$Na$_{0.31}$ \cite{NapoLiNa}, would therefore provide reliability to the viscoelastic model.

The static structure and the frequency moments, which are a required input of 
the viscoelastic model for the calculation of the dynamic properties, have 
been computed using the variational modified hypernetted chain (VMHNC) 
approximation \cite{VMHNC} which reproduces rather well the simulation 
results, although small differences can and do appear in the structural 
functions, which show up, for instance, in a small mismatch of the initial 
theoretical and simulation values of the ISFs.
While it would be possible to use the simulated $S_{ij}(k)$ instead of 
the VMHNC ones, the calculation by simulation of the fourth frequency 
moments is rather unreliable.

We recall that the viscoelastic model is expected to be accurate for systems
with $\gamma\sim 1$. Generalized hydrodynamics calculations of $\gamma$ have
been performed for liquid Li \cite{Canales}, liquid Pb \cite{Bryk01}, and 
liquid Na-K and K-Cs alloys \cite{Andrij_thesis}, leading in all cases to 
$\gamma$ values not far from 1. These results give some support to the
application of the viscoelastic model to liquid metals and alloys, in contrast
to other systems like Lennard Jones liquids \cite{Schepper88,Bryk97}
where $\gamma$ is larger.

\subsection{Model 1. Pseudo-binary system.}

Due to the particular concentration $x_1=x_2=0.5$, an additional 
symmetry appears in the system, that
implies that $S_{11}(k)=S_{22}(k)$, $F_{11}(k,t)=F_{22}(k,t)$ and so on.

\begin{figure}[h]
\includegraphics[scale=0.35,clip]{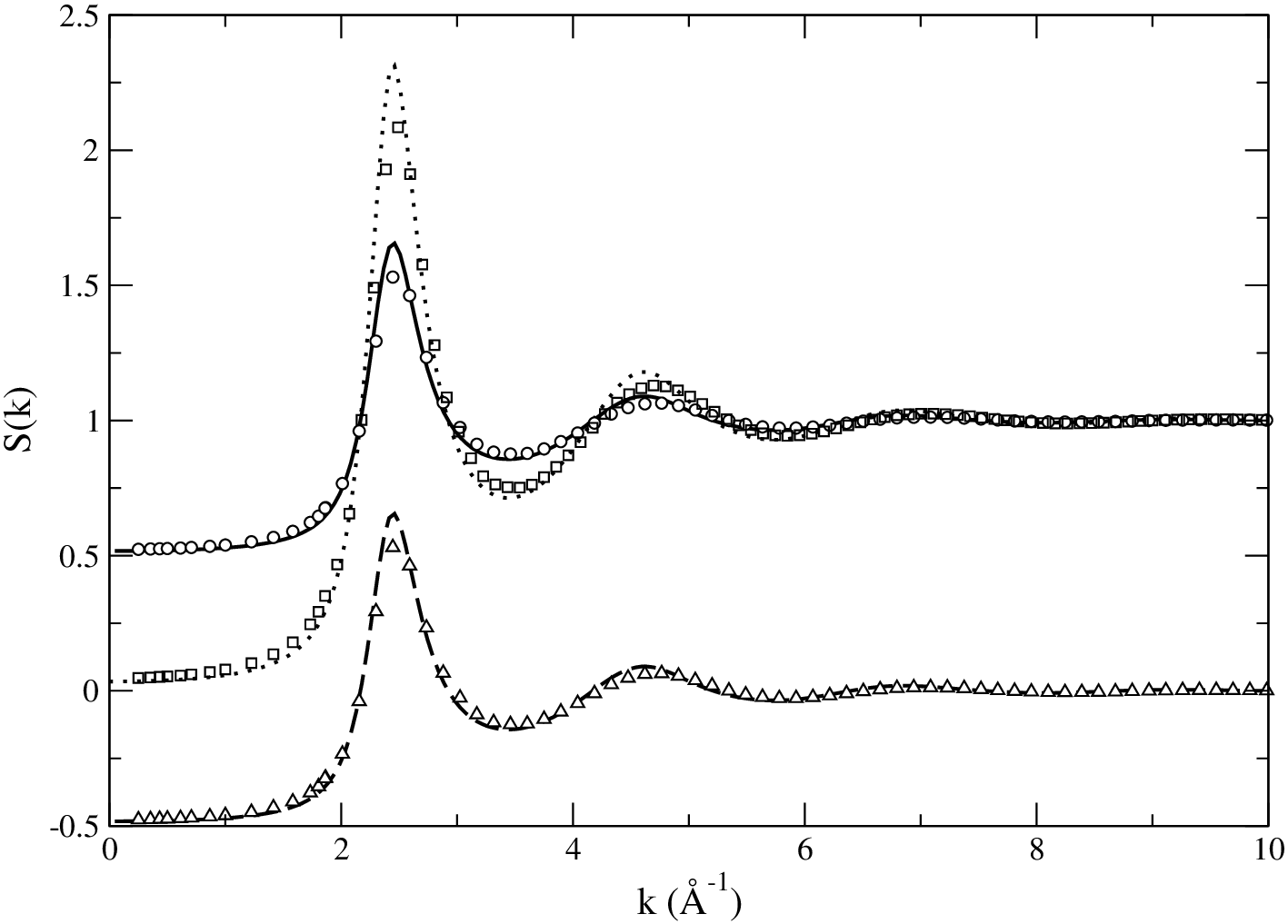}
\caption{Static partial structure factors for pseudobinary Li. Full line and circles: $S_{11}(k)=S_{22}(k)$, dashed line and triangles: $S_{12}(k)$, dotted line: one-component system $S(k)$. The lines are theoretical results and the symbols denote MD simulation results. \label{sijk_Li}} 
\end{figure}

\begin{figure}[h]
\includegraphics[scale=0.35,clip]{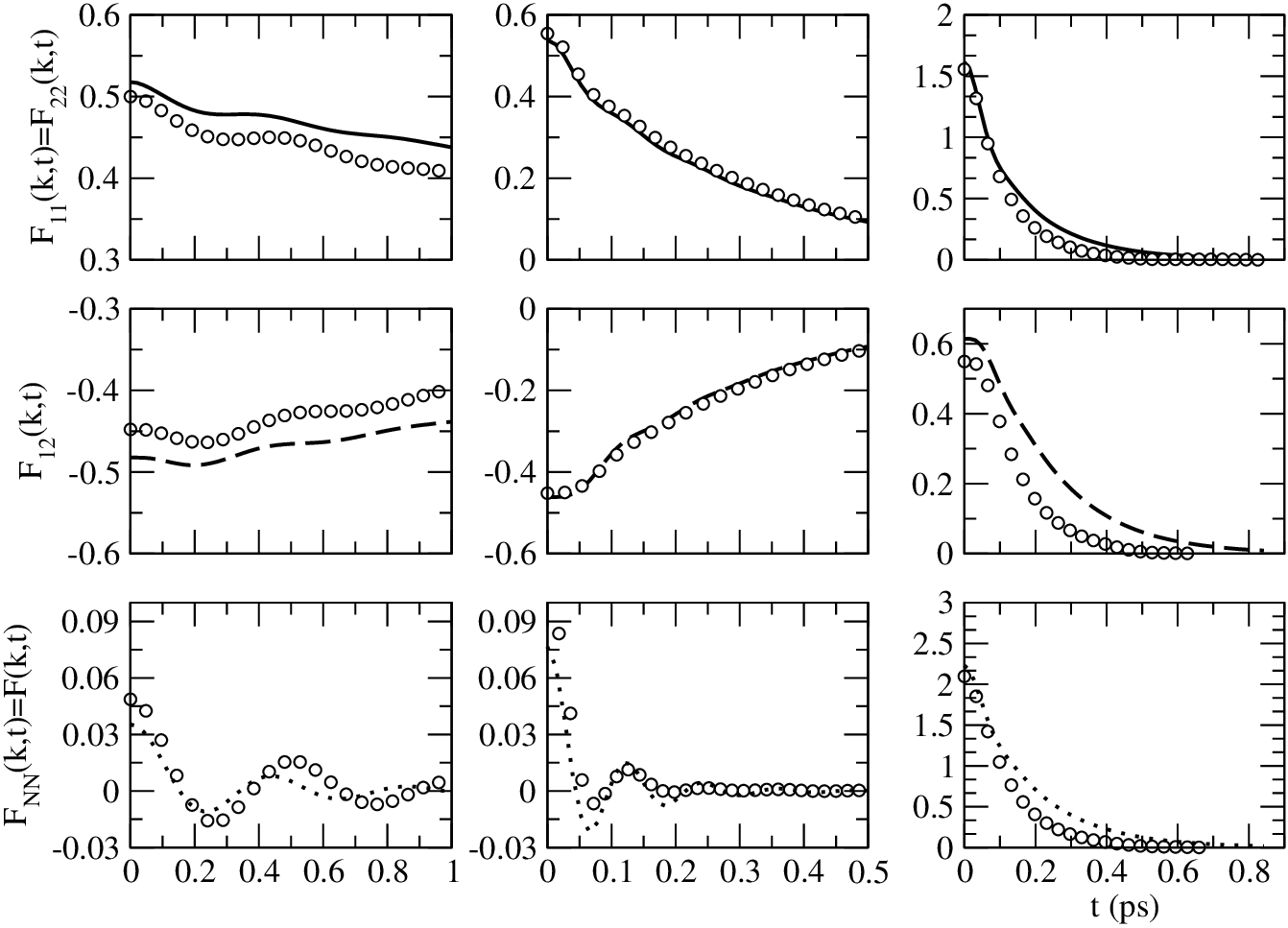}
\caption{Partial ISFs for pseudobinary Li. Full lines (upper panel): $F_{11}(k,t)=F_{22}(k,t)$, dashed line (mid panel): $F_{12}(k,t)$, dotted line (lower panel): $F_{NN}(k,t)$, which coincides with the one-component system $F(k,t)$. The symbols are the corresponding MD results.
Left column: $k=0.23$ \AA$^{-1}$, mid column: $k=1.23$ \AA$^{-1}$, right column: $k=2.51$ \AA$^{-1}$.
\label{fijkt_Li}} 
\end{figure}

Figure \ref{sijk_Li} shows the partial structure factors obtained form the simulation and the VMHNC theory, which are practically coincident. The first thing to note is that, as stated above, the small $k$ values attained by the $S_{ij}(k)$ are in magnitude close to $0.5$ (the concentrations) which is one order of magnitude larger than
the structure factor of the corresponding one-component system in the same region. This will imply a much larger initial value of the partial ISFs of the system as compared to the ISF of the one-component system.
The $F_{ij}(k,t)$ are shown in fig.~\ref{fijkt_Li}, where we have plotted both the MD results and the viscoelastic ones. The results for small $k$ are markedly different from the $F(k,t)$ of the one-component system, and not only in the initial value. The latter shows clear oscillations around zero, whereas the partial ISFs of the pseudobinary system show a large diffusive component which decays very slowly.

\begin{figure}[h]
\includegraphics[scale=0.35,clip]{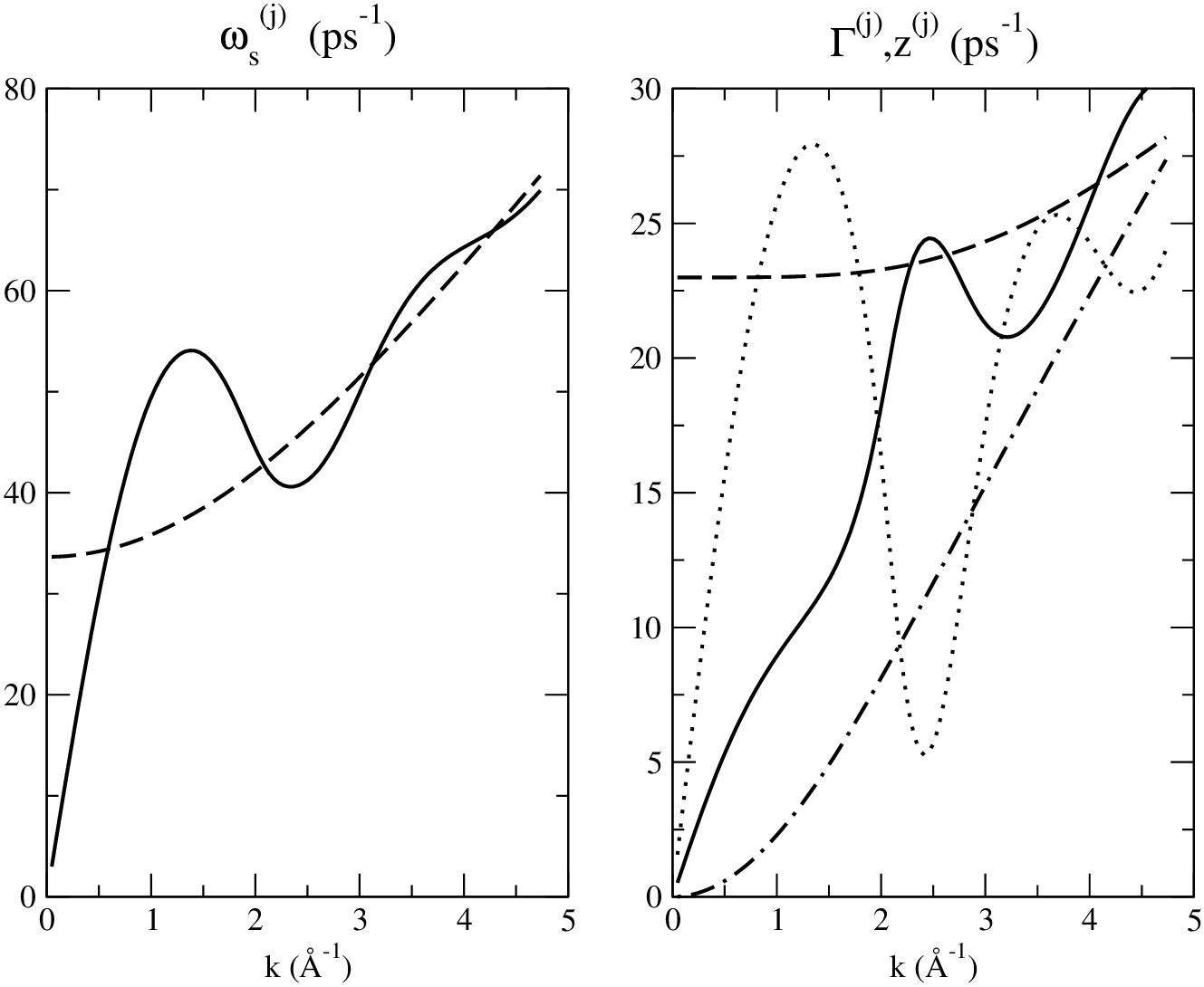}
\caption{Roots of pseudobinary Li in the viscoelastic model. Full lines: $\omega_s^{(1)}$ and $\Gamma^{(1)}$, dashed lines: $\omega_s^{(2)}$ and $\Gamma^{(2)}$, dotted line: $z^{(3)}$, dash-dotted line: $z^{(4)}$. \label{roots_Li}}
\end{figure}

\begin{figure}[h]
\includegraphics[scale=0.3,clip]{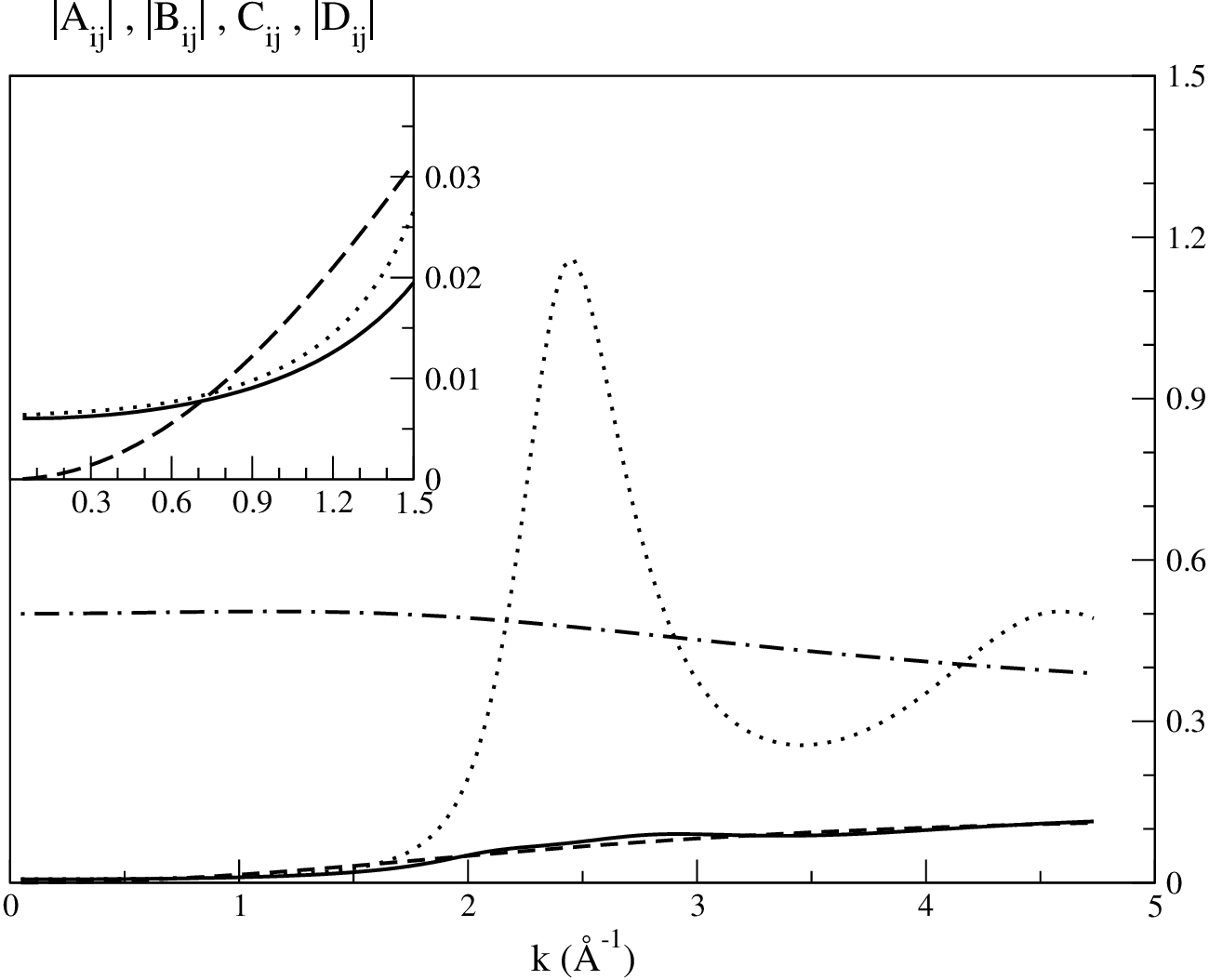}
\caption{Amplitudes of the different modes of pseudobinary Li in the viscoelastic model. Full lines: $A_{11}$, dashed lines: $B_{11}$, dotted line: $C_{11}$, dash-dotted line: $D_{11}$. \label{amps_Li}}
\end{figure}

The reason for this behavior can be found in the roots and corresponding amplitudes that are obtained within the viscoelastic model. The roots are plotted in fig.~\ref{roots_Li}, where we see that one of the propagating roots vanishes in the $k=0$ limit, while the other remains non-zero in this limit. We recall that in a pseudobinary system the determinant that leads to the roots factors into two third degree polynomials, which correspond to the modes of the DSF and the SDFS of the one-component system respectively. The first propagating root corresponds to the propagating mode in the DSF of the one-component system, while the non-vanishing root is an unexpected propagating mode that appears in the SDSF of the one-component system. Of the remaining two real roots, the largest one in magnitude for small $k$ is the diffusive mode of the DSF of the one-component system, whereas the smallest one in magnitude  corresponds to the diffusive mode of the SDSF of the one-component system. It is precisely this last root that is responsible for the behavior of the partial $F_{ij}(k,t)$, since its contribution is the most slowly decaying one. Moreover its amplitude is the largest for small $k$, as can be observed in fig.~\ref{amps_Li}. In fact we have only plotted the amplitudes corresponding to the 11 partial, since due to the symmetry imposed by the concentrations the amplitudes corresponding to the 22 partial are the same as the 11 ones, while we also have that $A_{12}=A_{11}, B_{12}=-B_{11}, C_{12}=C_{11}, D_{12}=-D_{11}$.
Note that $B_{11}$ goes to 0 as $k\to 0$, which is another characteristic feature of the kinetic modes. It is also interesting to observe the similarity between $C_{11}$ and the structure factors.
We can therefore conclude that the extremely slow decay of $F_{ij}(k,t)$ is a direct consequence of the influence of the single-particle dynamics on the behavior of the ISFs, and this is due to the very definition of the AL partial ISFs. When one comes to the BT partials no such problems arise, since there is a complete decoupling of the single particle dynamics, which goes exclusively into the concentration-concentration partial, and the collective dynamics, which is exclusively responsible for the number-number partial. Obviously this complete decoupling is due to the ideal character of this system.

\begin{figure}[h]
\includegraphics[scale=0.4,clip]{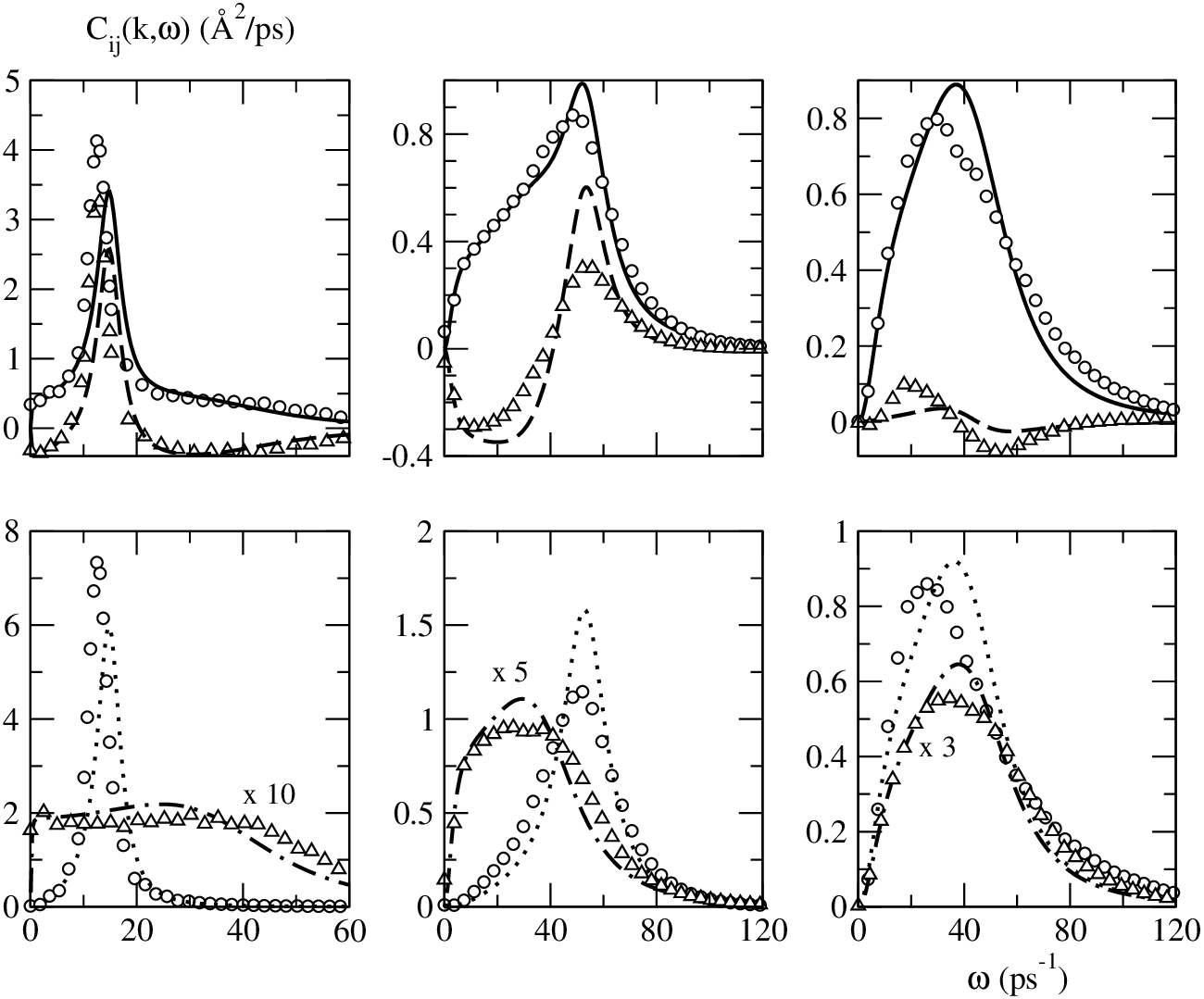}
\caption{Partial currents for pseudobinary Li. Lines are theoretical results and symbols denote simulation data. Full line and circles (upper panel): $C_{11}(k,\omega)=C_{22}(k,\omega)$, dashed line and triangles (upper panel): $C_{12}(k,\omega)$, dotted line and circles (lower panel): $C_{NN}(k,\omega)$, which coincides with the one-component system $C(k,\omega)$, dash-dotted line and triangles (lower panel): scaled $C_{cc}(k,\omega)$, which coincides with the one-component system $C_s(k,\omega)$.
Left column: $k=0.23$ \AA$^{-1}$, mid column: $k=1.23$ \AA$^{-1}$, right column: $k=2.18$ \AA$^{-1}$.
\label{cijkw_Li}} 
\end{figure}

The Fourier transforms of the longitudinal current correlation functions are directly related to the DSFs, through the relation $C_{ij}(k,\omega)=\frac{\omega^2}{k^2}S_{ij}(k,\omega)$ (and the same relation for the BT currents), while they can be calculated independently in the MD runs. They are especially useful in the cases where a slowly decaying term appears
in the ISFs that complicates the computation of the DSFs, while the multiplication by $\omega^2$ depletes the low frequency modes and enhances the high frequency ones. In fig.~\ref{cijkw_Li} we show the functions $C_{ij}(k,\omega)$ and also the BT ones. We always find a clear peak and a minimum in $C_{12}(k,\omega)$, while there is a clear maximum in $C_{11}(k,\omega)$ at the position of the peak of $C_{12}(k,\omega)$, and sometimes a shoulder at the position of the minimum. These are of course the remnants of the two propagating modes. The coincidence of the shoulder of $C_{11}(k,\omega)$ with the
minimum of $C_{12}(k,\omega)$ suggests that the second propagating mode on top of being kinetic is optical in character, with the two ``species" moving in opposite directions.
The appearance of an optical mode in a pseudobinary liquid system resembles a
similar effect that appears for crystals: 
when a linear chain with equilibrium distance $a$ 
is considered as a chain with equilibrium distance $2a$ with a basis formed by
two ``different" atoms separated by a distance $a$ \cite{Ash_book},
then the first Brillouin zone becomes half the original one, and the original
dispersion relation becomes folded into the new one leading to the optical
mode.
The BT currents on the other hand always show a single peak, 
in the $NN$ case (which is the current of the one-component system) at 
the positions of the peaks in the 11 and 12 functions, and in the $cc$ case 
(which is the self-current of the one-component system) at the position of the minimum of the 12 function.

The appearance of the peak in the self-currents of the one-component system is very clear in the MD simulations, showing that also in the single-particle properties of this system there are indeed propagating contributions, of kinetic character. These modes had not been reported previously, at least to our knowledge, since the focus has usually been put on the initial value and the width of the SDSF of the systems, analyzing the behavior of these properties within the different theoretical approaches.
The reproduction of these propagating modes by the viscoelastic model (for the single-particle properties)
is therefore an interesting property of this theory, and warrants a wider application of the model
in the analysis of the single-particle properties of, at least, this type of dense liquids. 

\subsection{Li-based alloys: ideal and compound forming mixtures.}

\begin{figure}[h]
\includegraphics[scale=0.35,clip]{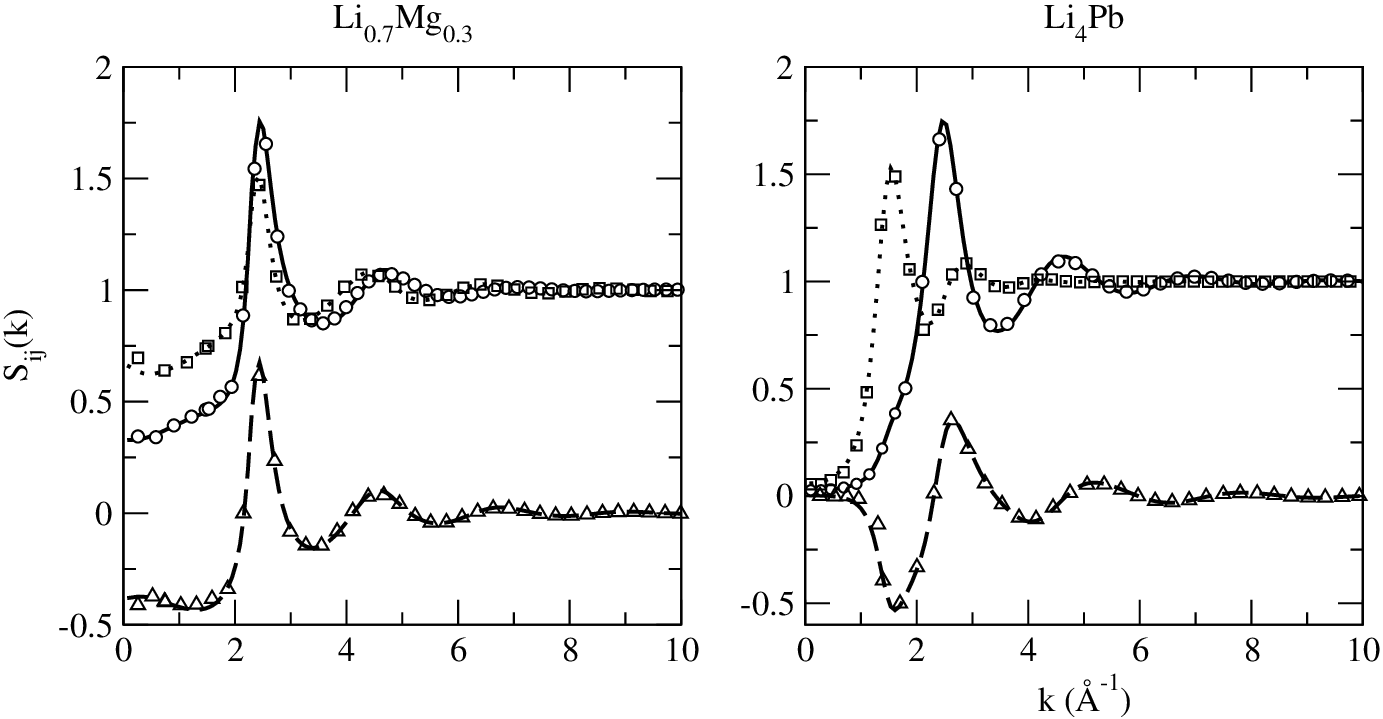}
\caption{Static partial structure factors for \LiMg and Li$_4$Pb. Full lines and circles: $S_{11}(k)$, dotted lines and squares: $S_{22}(k)$, dashed lines and triangles: $S_{12}(k)$. The lines are theoretical results and the symbols denote MD simulation results.\label{sijLiMgPb}} 
\end{figure}

\begin{figure}[h]
\includegraphics[scale=0.35,clip]{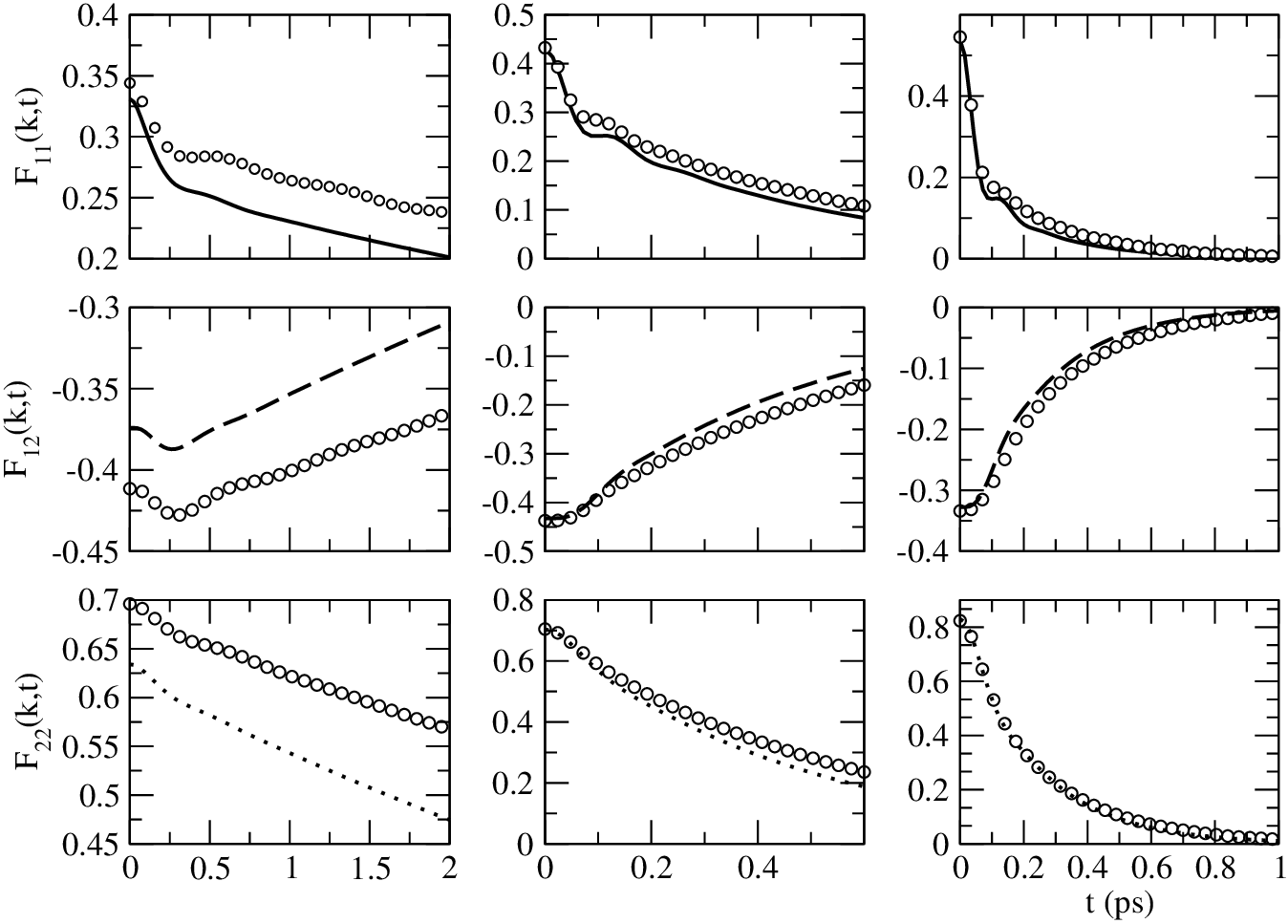}
\caption{Partial ISFs for Li$_{0.7}$Mg$_{0.3}$.
Lines are theoretical results and symbols are simulation data. 
Full lines (upper panel): $F_{11}(k,t)$, 
dashed line (mid panel):  $F_{12}(k,t)$, 
dotted line (lower panel): $F_{22}(k,t)$.
Left column: $k=0.28$ \AA$^{-1}$ (theory), $0.26$ \AA$^{-1}$ (simulation);
 mid column: $k=1.25$ \AA$^{-1}$ (theory), $1.23$ \AA$^{-1}$ (simulation);
right column: $k=1.88$ \AA$^{-1}$ (theory and simulation).
\label{fijkt_LiMg}} 
\end{figure}

\begin{figure}[h]
\includegraphics[scale=0.35,clip]{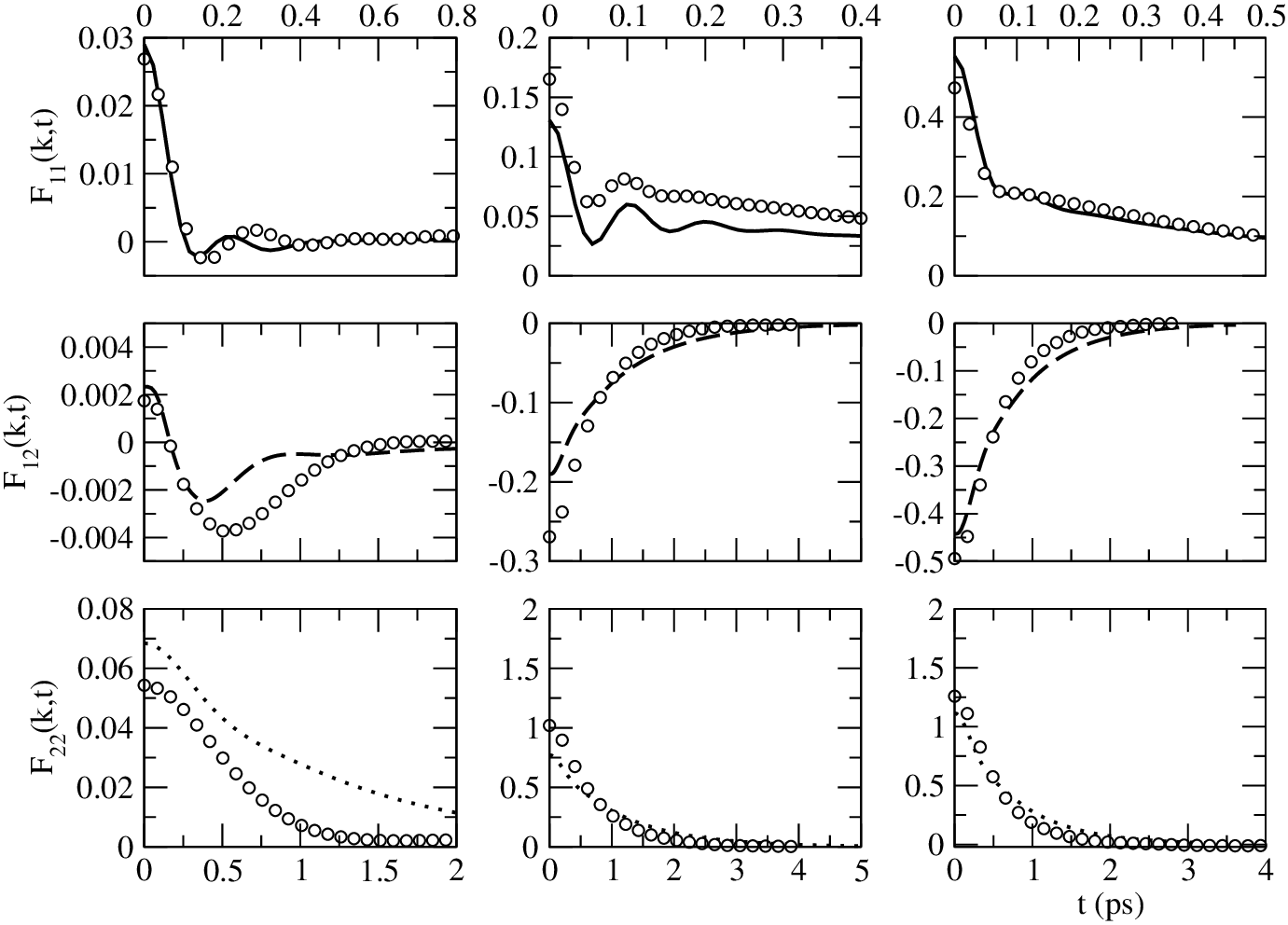}
\caption{Partial ISFs for Li$_4$Pb. 
The meaning of the lines and symbols are the same as in the 
previous figure.
Left column: $k=0.31$ \AA$^{-1}$ (theory), $0.27$ \AA$^{-1}$ (simulation);
 mid column: $k=1.30$ \AA$^{-1}$ (theory and simulation);
right column: $k=1.76$ \AA$^{-1}$ (theory), $1.74$ \AA$^{-1}$ (simulation).
Note the different scales in the time axis for $F_{11}(k,t)$.
\label{fijkt_LiPb}} 
\end{figure}

In the following we will denote Li as component 1. The partial static structure factors of both liquid alloys are shown in fig.~\ref{sijLiMgPb}. Those corresponding to Li$_{0.7}$Mg$_{0.3}$ are similar in character to those of the pseudobinary alloy, except for the difference in concentration, so the values of $S_{ij}(k=0)$ are larger than that of a one-component system. 
On the other hand the results for Li$_4$Pb are very different, in particular the values of $S_{ij}(k=0)$ are all the three similar in magnitude to the case of one-component system; 
moreover the dip in $S_{12}(k)$ and the coincident positions of the maximum of $S_{11}(k)$ and the minimum of $S_{22}(k)$ is indicative of a kind of ionic ordering in the melt \cite{NapoPRE64}. 
The corresponding ISFs are also strikingly different for the two systems, especially for small $k$, as observed in figs.~\ref{fijkt_LiMg} and \ref{fijkt_LiPb}. As in the pseudobinary system, the ISFs of Li$_{0.7}$Mg$_{0.3}$ have a very
 slowly decaying diffusive term for small $k$. In the case of Li$_4$Pb, the slowly decaying term is practically absent in $F_{11}(k,t)$ and while there is indeed such a term in $F_{12}(k,t)$ and especially in $F_{22}(k,t)$, it decays much more rapidly than in \LiMg (observe the $y$-axis scales in figs.~\ref{fijkt_LiMg} and \ref{fijkt_LiPb}) even though the Pb ionic mass is much larger than the Mg one.

\begin{figure}[h]
\includegraphics[scale=0.3,clip]{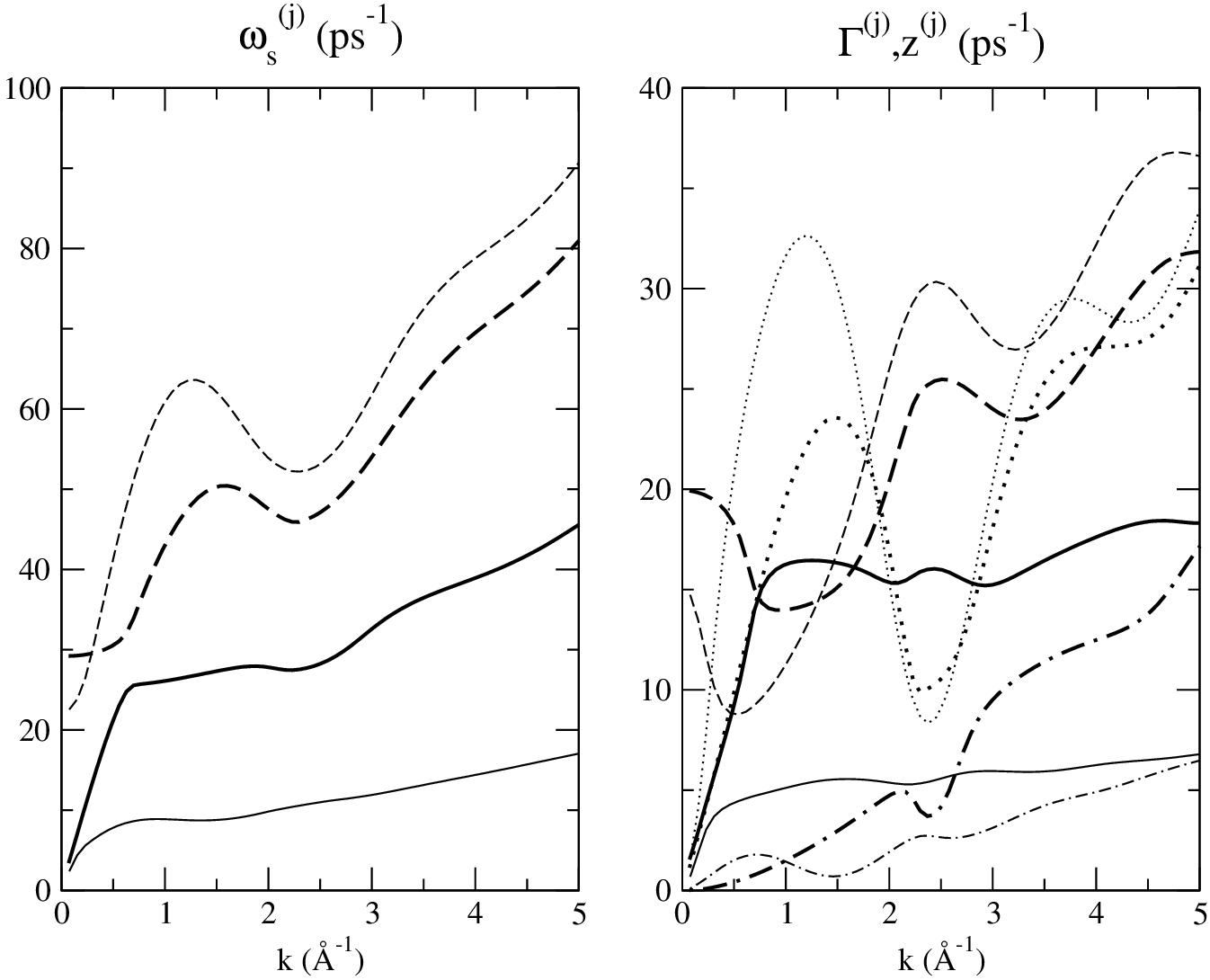}
\caption{Roots of \LiMg (thick lines) and Li$_4$Pb (thin lines) in the viscoelastic model. Full lines: $\omega_s^{(1)}$ and $\Gamma^{(1)}$, dashed lines: $\omega_s^{(2)}$ and $\Gamma^{(2)}$, dotted line: $z^{(3)}$, dash-dotted line: $z^{(4)}$.  \label{roots_LiMgPb}}
\end{figure}

Again the reason for such behaviors can be traced back to the roots and the amplitudes of the modes associated to the three partials. 
The roots are shown in fig.~\ref{roots_LiMgPb}. We see that, due to the mass difference, the frequency $\omega_s^{(1)}$ and damping $\Gamma^{(1)}$ of the low frequency root are larger for \LiMg than for Li$_4$Pb, while the high frequency root and the larger real root have quite similar magnitude and variation with $k$ for the two systems, being much more influenced by the masses of the components and the thermodynamic state than by the particular structure of the alloy. The smallest real root, $z^{(4)}$, 
also behaves different for \LiMg and Li$_4$Pb, staying much closer to $\Gamma^{(1)}$ for the latter. For small $k$, in particular,
$z^{(4)}$ is significantly larger for Li$_4$Pb than for \LiMg, which explains the faster decay of $F_{12}(k,t)$ and $F_{22}(k,t)$ in Li$_4$Pb. 

\begin{figure}[h]
\includegraphics[scale=0.3,clip]{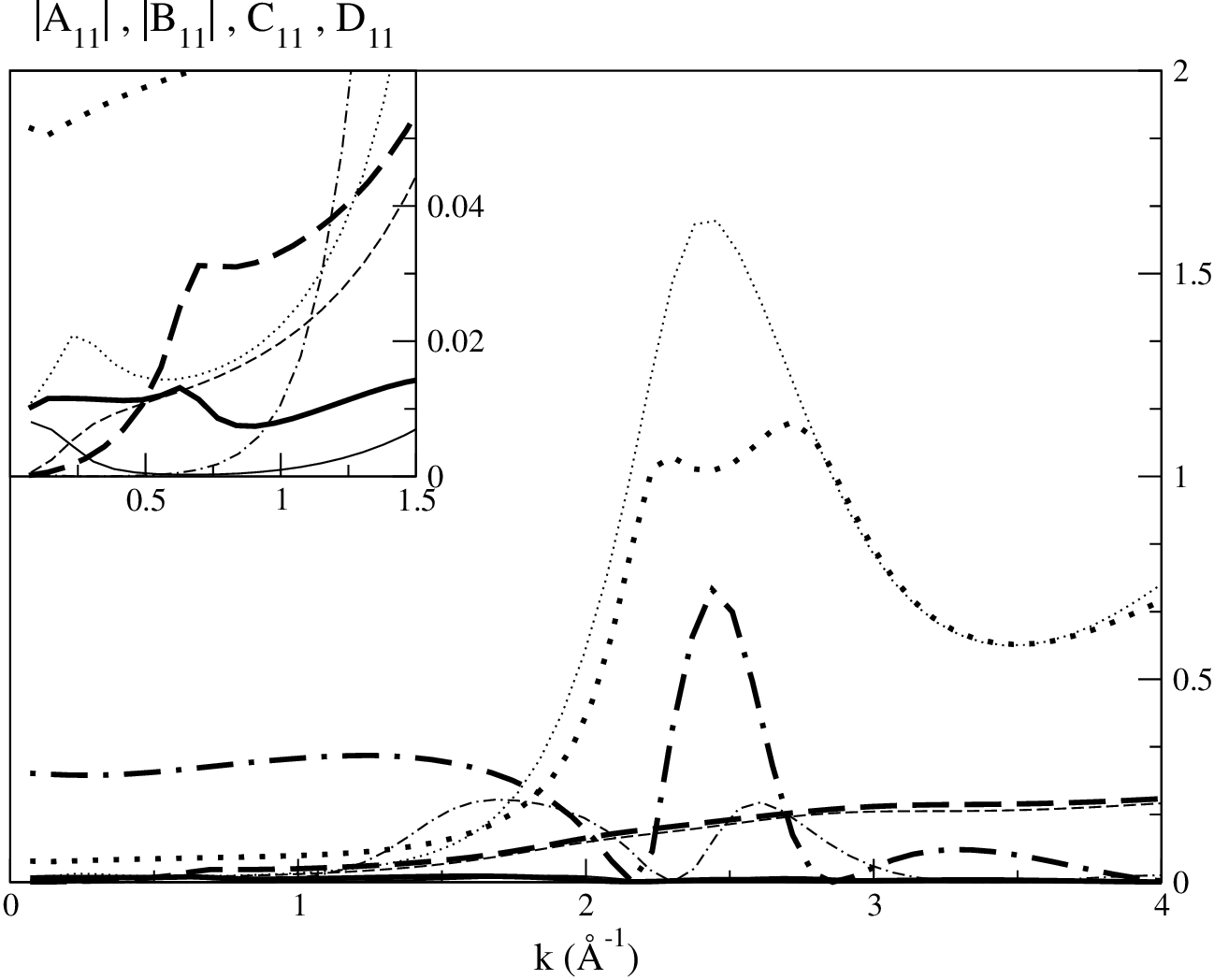}
\caption{Amplitudes of the different modes of $F_{11}(k,t)$ for \LiMg (thick lines) and Li$_4$Pb (thin lines) in the viscoelastic model. Full lines: $A_{11}$, dashed lines: $B_{11}$, dotted line: $C_{11}$, dash-dotted line: $D_{11}$. \label{amps11LiMgPb}}
\end{figure}

\begin{figure}[h]
\includegraphics[scale=0.3,clip]{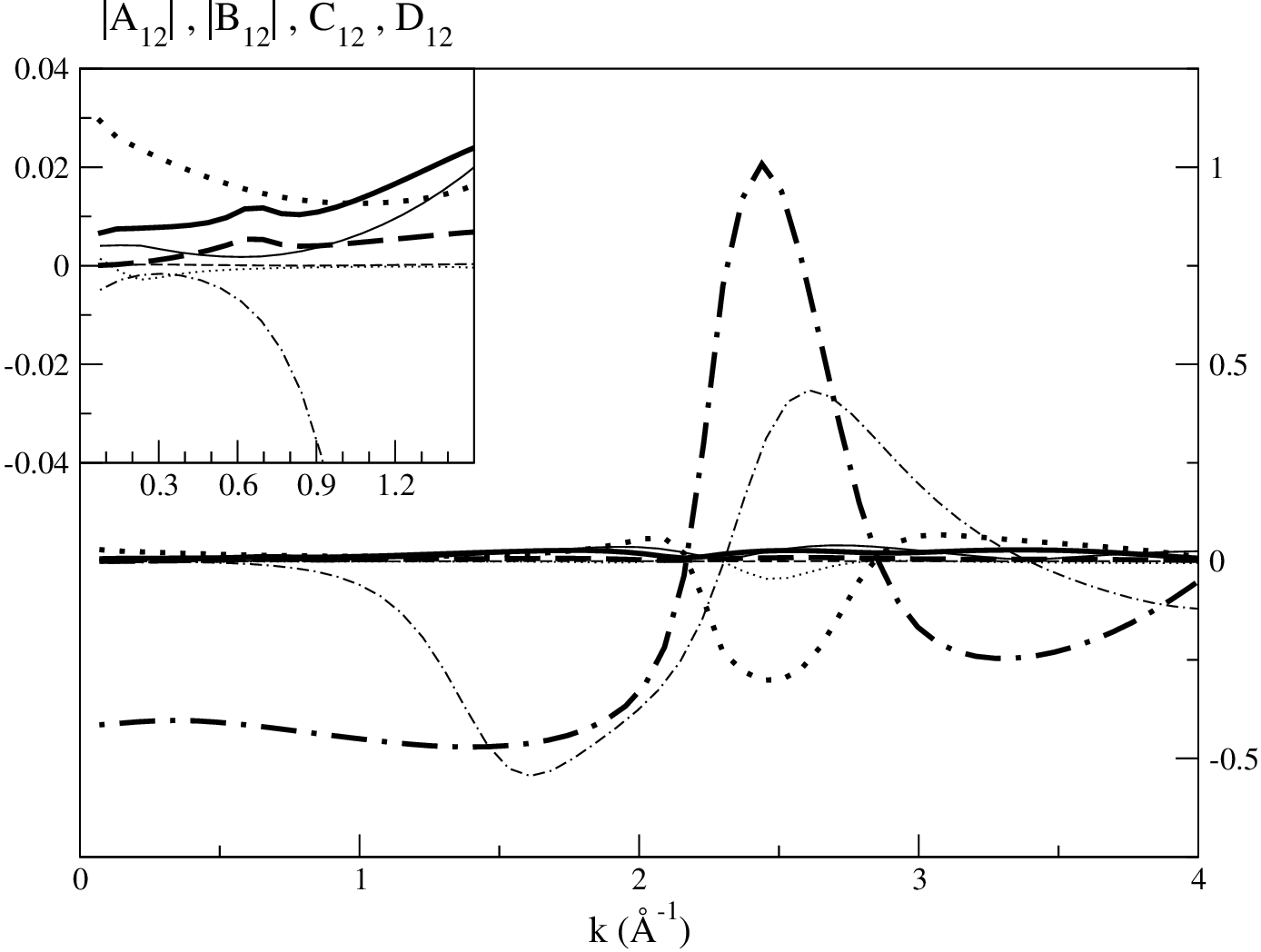}
\caption{Amplitudes of the different modes of $F_{12}(k,t)$ for \LiMg (thick lines) and Li$_4$Pb (thin lines) in the viscoelastic model. Full lines: $A_{12}$, dashed lines: $B_{12}$, dotted line: $C_{12}$, dash-dotted line: $D_{12}$. \label{amps12LiMgPb}}
\end{figure}

\begin{figure}[h]
\includegraphics[scale=0.3,clip]{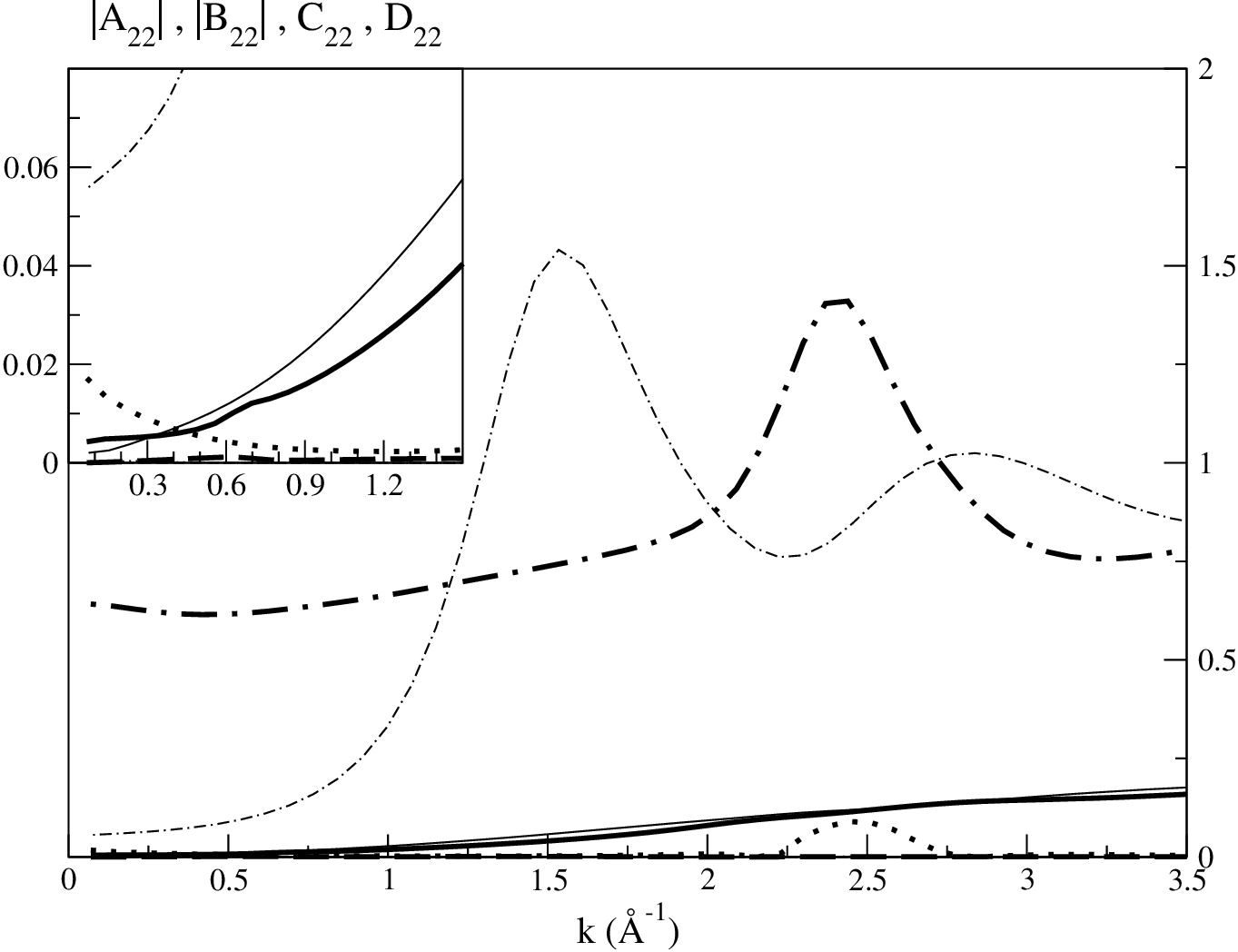}
\caption{Amplitudes of the different modes of $F_{22}(k,t)$ for \LiMg (thick lines) and Li$_4$Pb (thin lines) in the viscoelastic model. Full lines: $A_{22}$, dashed lines: $B_{22}$, dotted line: $C_{22}$, dash-dotted line: $D_{22}$. \label{amps22LiMgPb}}
\end{figure}

Figs.~\ref{amps11LiMgPb}-\ref{amps22LiMgPb} show the amplitudes for 
$F_{ij}(k,t)$. The first noticeable feature is that the $F_{22}(k,t)$ 
functions are completely dominated for all $k$ by the smallest real root, and to a lesser extent by the low frequency root, the other two amplitudes being smaller, except in the case of \LiMg in the region of the main peak of the corresponding structure factor and  for small $k$, where the amplitude corresponding to the larger real root is also significant.
The case of the $F_{12}(k,t)$ functions is rather similar, with the difference that for small $k$ in \LiMg the three amplitudes corresponding to the two propagating roots and the larger real root are of similar magnitude.
Note again the striking similarity between the amplitudes of the smallest real root and the structure factors, with $D_{12}$ and $D_{22}$ following respectively $S_{12}(k)$ and $S_{22}(k)$. The different behavior of these structure factors for the two systems implies now that the amplitude of the diffusive mode is much smaller 
for Li$_4$Pb than for Li$_{0.7}$Mg$_{0.3}$.
As for the $F_{11}(k,t)$ functions, the structure of the amplitudes is 
richer. 
For medium and large $k$ the dominant term is $z^{(3)}$, the larger real root, with its amplitude $C_{11}$ following $S_{11}(k)$. For small $k$, on the other hand, the situation is different; $D_{11}$ is dominant for \LiMg, while for Li$_4$Pb it is still $C_{11}$ the dominant contribution, with an intermediate region around 1.5 \AA$^{-1}$, where $D_{11}$ is of similar magnitude. Therefore the overall decay of $F_{11}(k,t)$ is dictated by $z^{(3)}$, which being much larger than $z^{(4)}$ induces a much faster decay; note the smaller scale in the time axis for $F_{11}(k,t)$ in fig.~\ref{fijkt_LiPb}.

It is also interesting to observe that out of the two propagating roots, 
the dominant one for very small $k$ is the low frequency one, and therefore 
in this limit all the three partials oscillate with the same frequency,
as happens within the hydrodynamic model. 
There is however a transition from this behavior to the dominance of the 
high frequency mode for $k$ values around $0.2$ \AA$^{-1}$ for Li$_4$Pb 
and around $0.5$ \AA$^{-1}$ for \LiMg. This will have a direct impact on 
the presence of low and/or high frequency peaks in the partial DSFs.

\begin{figure}[h]
\includegraphics[scale=0.3,clip]{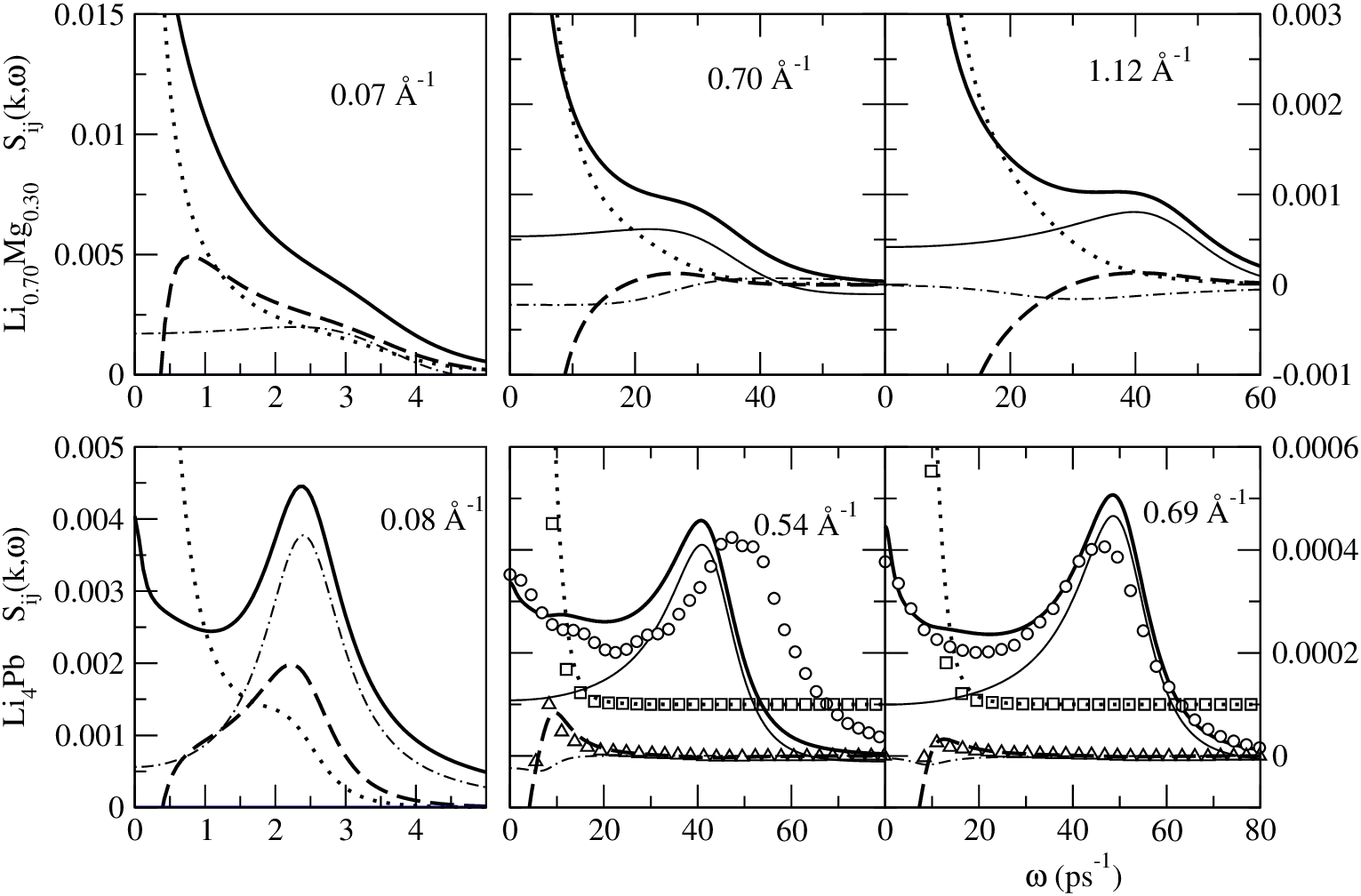}
\caption{Dynamic structure factors for \LiMg (upper pannel) and Li$_4$Pb (lower pannel) at wavevectors
shown in the graphs. Full lines: $S_{11}(k,\omega)$, dashed lines: $S_{12}(k,\omega)$, dotted line: $S_{22}(k,\omega)$. The thin dash-dotted line denotes the low frequency contribution to $S_{11}(k,\omega)$
and the thin full line is the corresponding high frequency contribution, which is negligible for the lowest $k$. 
Symbols denote simulation data at wavevectors $0.53$, and $0.65$ \AA$^{-1}$ 
for Li$_4$Pb. For these graphs $S_{22}(k,\omega)$ have been shifted upwards 
by $0.0001$ to improve visibility.
\label{sijkw_LiMgPb}}
\end{figure}

In fig.~\ref{sijkw_LiMgPb} we show the DSFs for both alloys at a small wavevector, a wavevector in the transition region and at a larger $k$. 
In the case of Li$_4$Pb we observe at the smallest $k$ a clear peak in 
$S_{11}(k,\omega)$ and $S_{12}(k,\omega)$ and a shoulder in the 
$S_{22}(k,\omega)$, all at the same frequency, akin to hydrodynamic sound 
propagation, even though the presence of the diffusive term masks the 
propagating mode and
$S_{22}(k,\omega)$ shows no side peak. For larger $k$ we find a prominent peak in $S_{11}(k,\omega)$ due to the high frequency contribution, while there is no indication of any peak in $S_{22}(k,\omega)$. This
appearance of high frequency peaks in the DSF of the light component (11) and its absence in the DSF of the heavy component (22) has been associated with the so called ``fast sound" phenomenon, which as we see appears naturally within the viscoelastic model as a consequence of the two propagating roots of the model.
In the case of \LiMg the diffusive contribution to $S_{ij}(k,\omega)$ is very large for small $k$, and completely covers the (low frequency) propagating contribution, so that no side peaks appear in the DSFs
even for rather small $k$ values. Anyhow this contribution is indeed present and shows up as a rather weak shoulder at the same frequency in all the three $S_{ij}(k,\omega)$. 
For larger $k$ the higher frequency component becomes more important in $S_{11}(k,\omega)$ and due to its larger frequency is less covered by the diffusive contribution, so that for sufficiently large $k$ a peak finally develops in $S_{11}(k,\omega)$, whereas $S_{22}(k,\omega)$, whose dominant propagating contribution is the low frequency one, never develops a side peak. Therefore the situation is rather similar to the case of Li$_4$Pb, except that the separation between the frequencies of the low and the high frequency modes is smaller due to the smaller mass of Mg, and that the magnitude of the diffusive mode in \LiMg definitely obscures the analysis.

\begin{figure}[h]
\includegraphics[scale=0.35,clip]{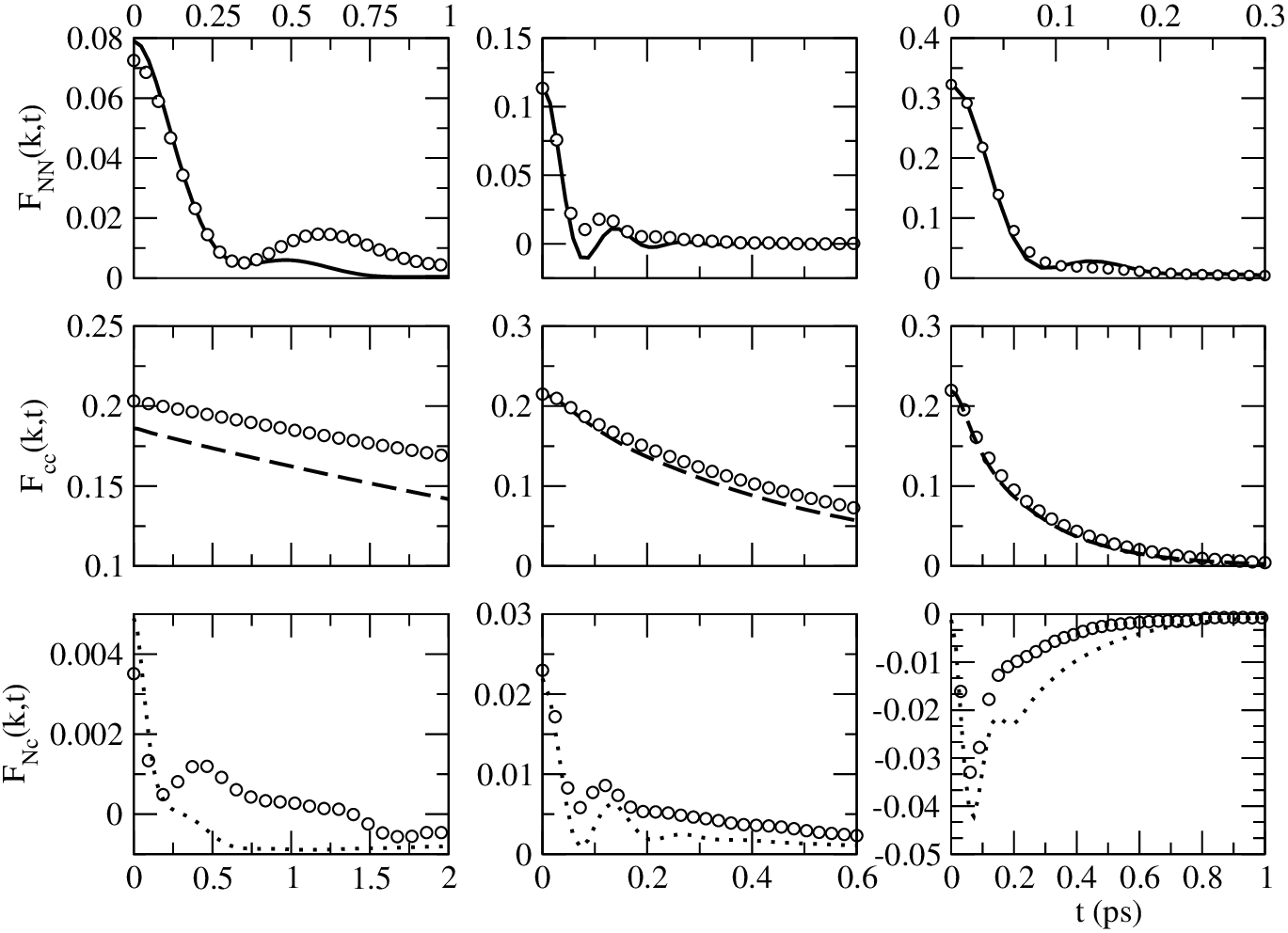}
\caption{Bhatia-Thornton ISFs for \LiMg. The $k$ values for the different columns are the same as in 
fig.~\ref{fijkt_LiMg}.
Note in some cases the different scales in the time axis for $F_{NN}(k,t)$.
Lines are theoretical results and symbols are simulation data.
\label{fbt_LiMg}} 
\end{figure}

\begin{figure}[h]
\includegraphics[scale=0.35,clip]{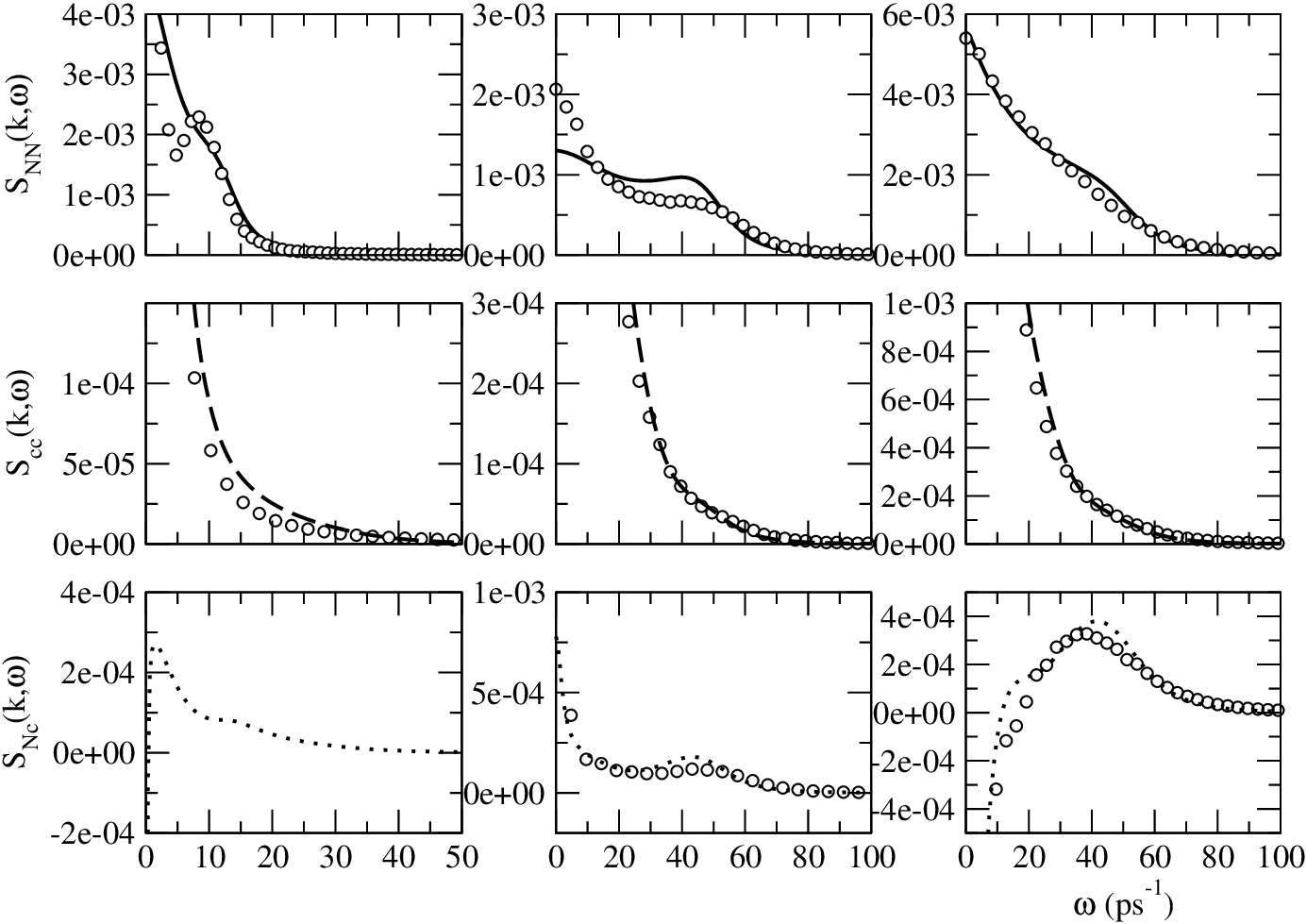}
\caption{Bhatia-Thornton DSFs for \LiMg. 
The $k$ values for the different columns are the same as in fig.~\ref{fijkt_LiMg}. Lines are theoretical results and symbols are simulation data.
\label{sbt_LiMg}} 
\end{figure}

\begin{figure}[h]
\includegraphics[scale=0.35,clip]{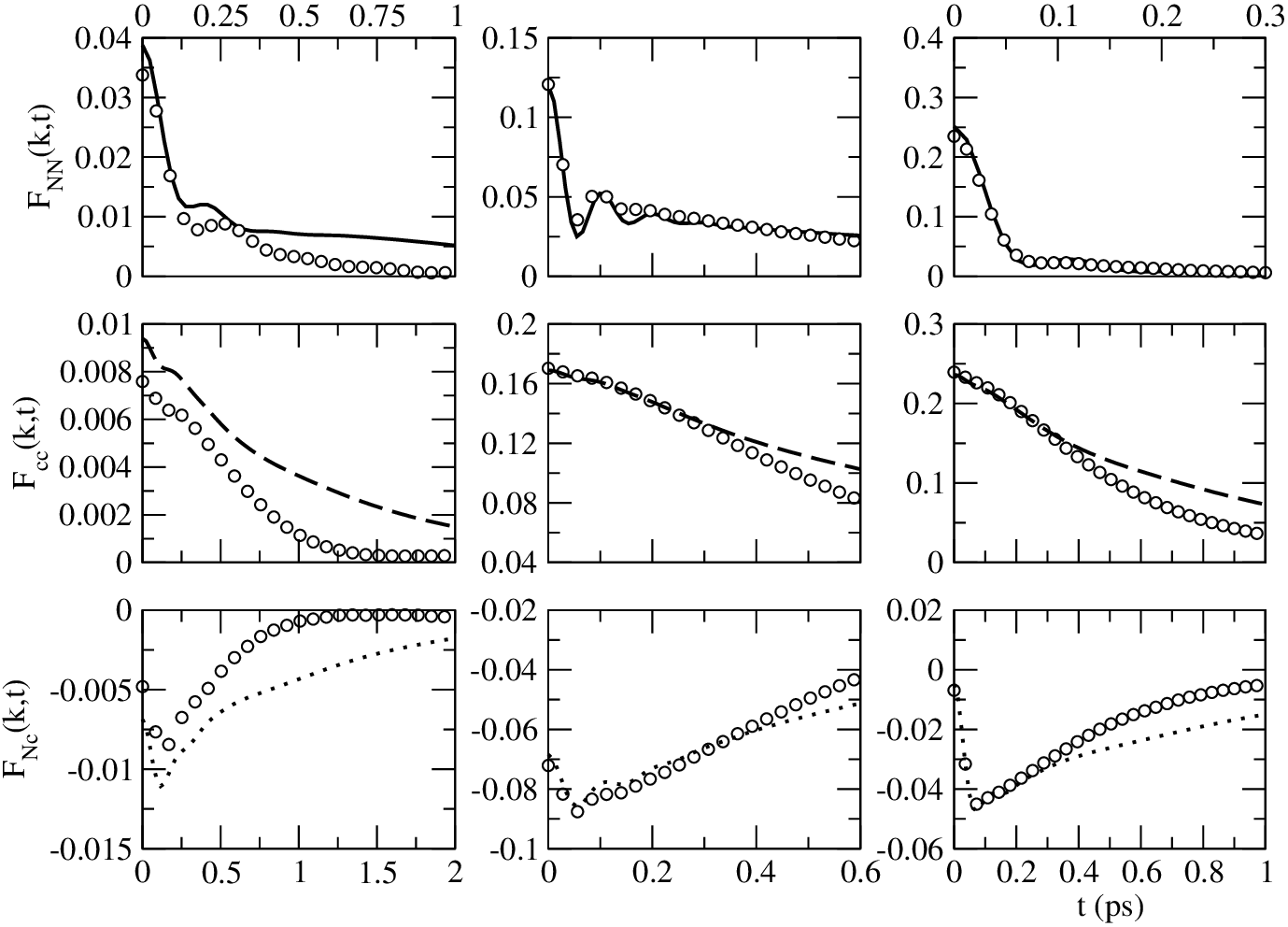}
\caption{Bhatia-Thornton ISFs for Li$_4$Pb. The $k$ values for the different columns are the same as in 
fig.~\ref{fijkt_LiPb}. Lines are theoretical results and symbols are simulation data.
Note in some cases the different scales in the time axis for $F_{NN}(k,t)$.
\label{fbt_LiPb}} 
\end{figure}

\begin{figure}[h]
\includegraphics[scale=0.35,clip]{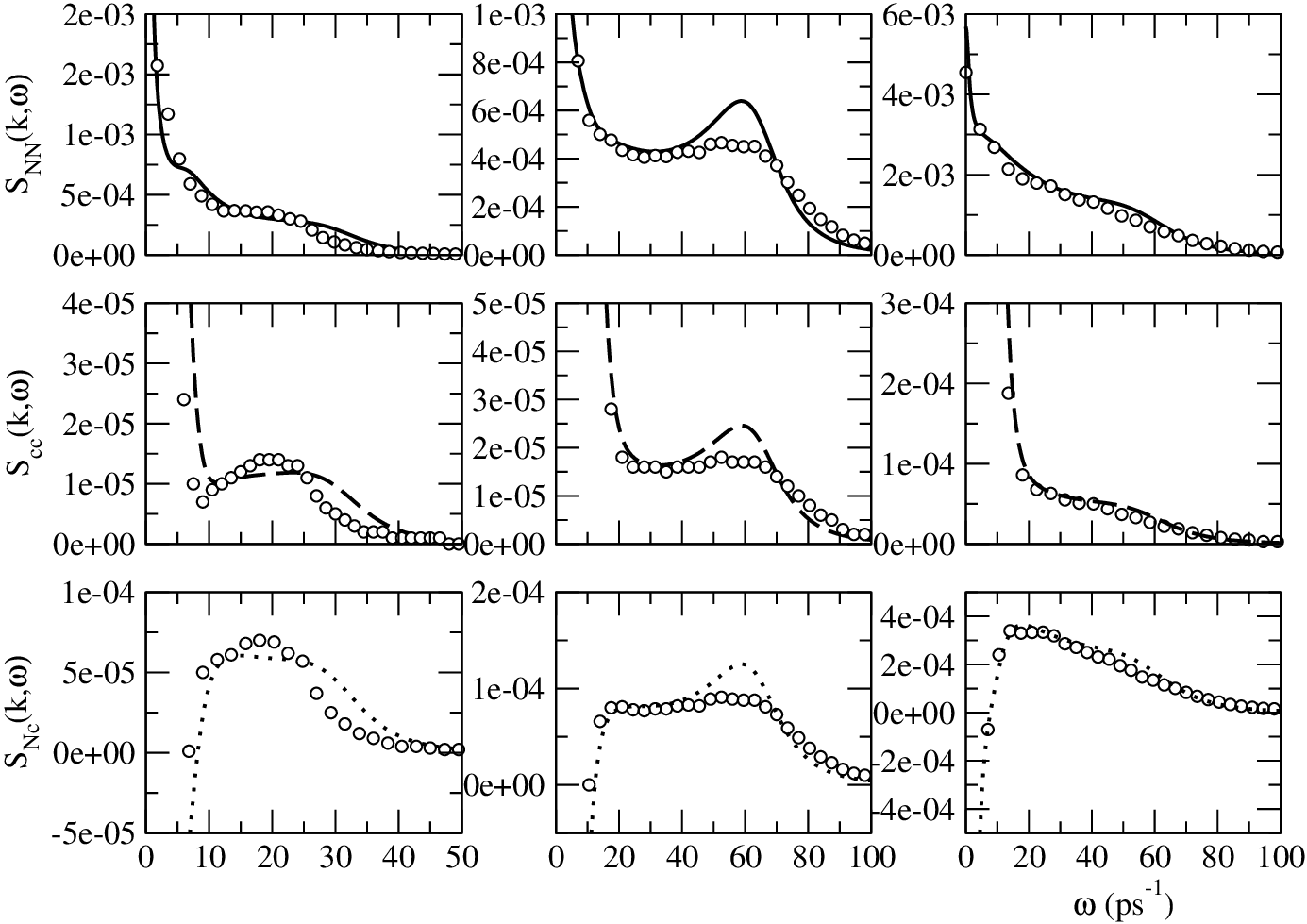}
\caption{Bhatia-Thornton DSFs for Li$_4$Pb. 
The $k$ values for the different columns are the same as in fig.~\ref{fijkt_LiPb}. Lines are theoretical results and symbols are simulation data.
\label{sbt_LiPb}} 
\end{figure}

Coming to the BT functions, one can expect that \LiMg, being a quasi-ideal system, should behave similar to the pseudobinary system, i.e., the single-particle dynamics should dominate the concentration-concentration partials, while the collective dynamics should dominate the number-number partials. 
This is indeed what happens, as shown in fig.~\ref{fbt_LiMg}, where we see that $F_{NN}(k,t)$ in fact decays much faster than the partials, which is similar 
to the behavior obtained for the pseudobinary system, 
$F_{cc}(k,t)$ is practically purely diffusive, as happens for the SISF of a one-component system, and $F_{Nc}(k,t)$ is very much smaller than the two other partials. Obviously this is also reflected in the dynamic structure factors, which are plotted in fig.~\ref{sbt_LiMg}. $S_{NN}(k,\omega)$ shows clear shoulders or peaks, while $S_{cc}(k,\omega)$ shows neither side peaks nor shoulders 
(note that for the smallest $k$ shown the MD $F_{Nc}(k,t)$ is too noisy to 
allow a reliable Fourier transform).
On the contrary, the case of Li$_4$Pb is different, as shown in figs.~\ref{fbt_LiPb} and \ref{sbt_LiPb}.
$F_{NN}(k,t)$ now exhibits a certain diffusive component and $F_{cc}(k,\omega)$ does show oscillations.
As a consequence, $S_{cc}(k,\omega)$ presents well defined side peaks or shoulders, indicative of 
propagating concentration modes. It is important to note that the magnitude of these peaks is very small,
around 10$^{-5}$, compared to the value of $S_{cc}(k,\omega=0)$, which is far outside the range of the graph, being around $10^{-3}-10^{-2}$ depending on the wavevector.
We stress that the reproduction of the different behaviors exhibited by the
two alloys, even at the level of such small details, represents a stringent 
test which is well fulfilled by the viscoelastic model.

\begin{figure}[h]
\includegraphics[scale=0.35,clip]{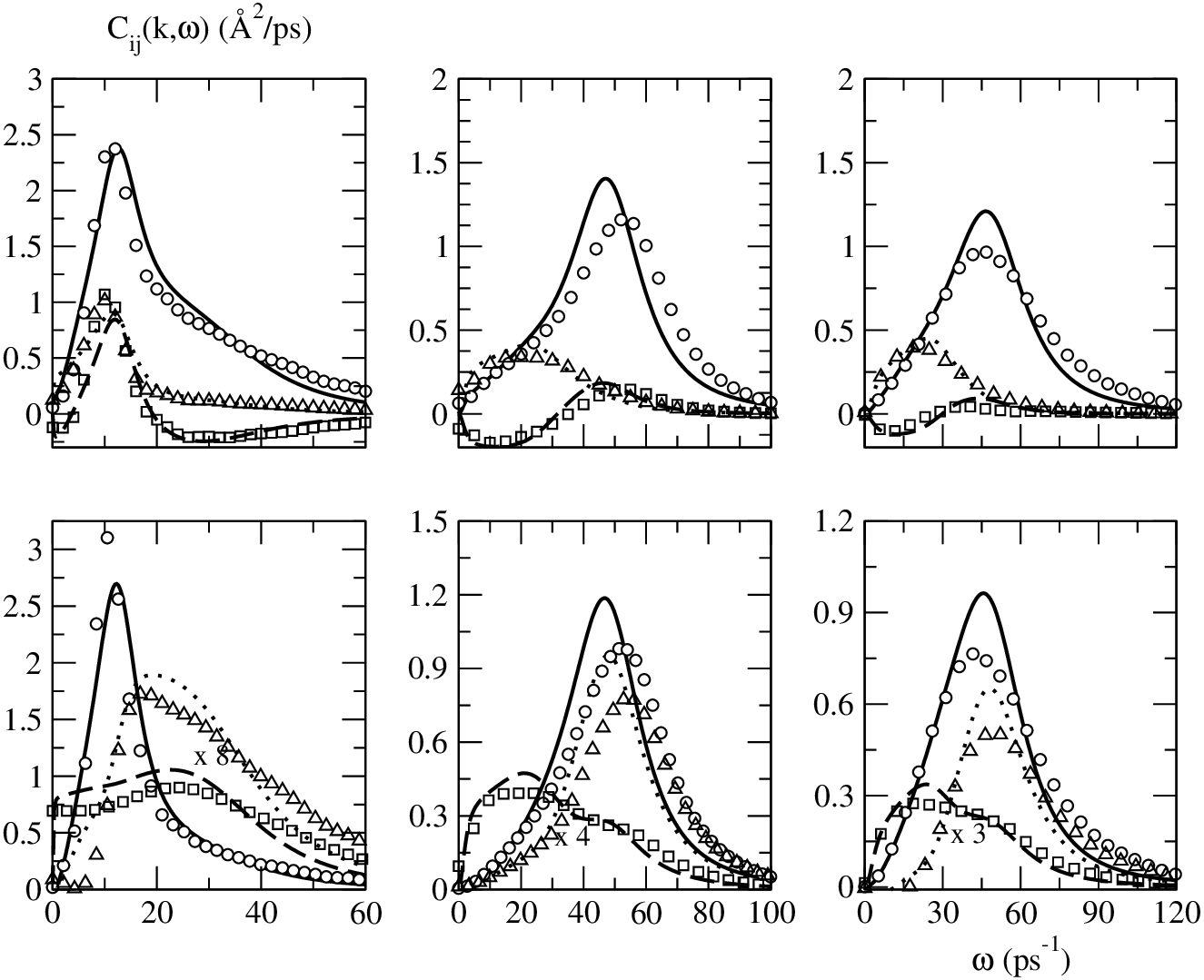}
\caption{Longitudinal current correlation functions for \LiMg. 
Lines are theoretical results and symbols are simulation data. 
Upper pannel: Ashcroft-Langreth partials; full line and circles: $C_{11}(k,\omega)$, dashed line and squares: $C_{12}(k,\omega)$, dotted line and triangles: $C_{22}(k,\omega)$
Lower pannel: Bhatia-Thornton partials; full line and circles: $C_{NN}(k,\omega)$, dashed line and squares: scaled $C_{cc}(k,\omega)$, dotted line and triangles: scaled $C_{Nc}(k,\omega)$.
The $k$ values for the different columns are the same as in fig.~\ref{fijkt_LiMg}.
\label{ckw_LiMg}} 
\end{figure}

\begin{figure}[h]
\includegraphics[scale=0.35,clip]{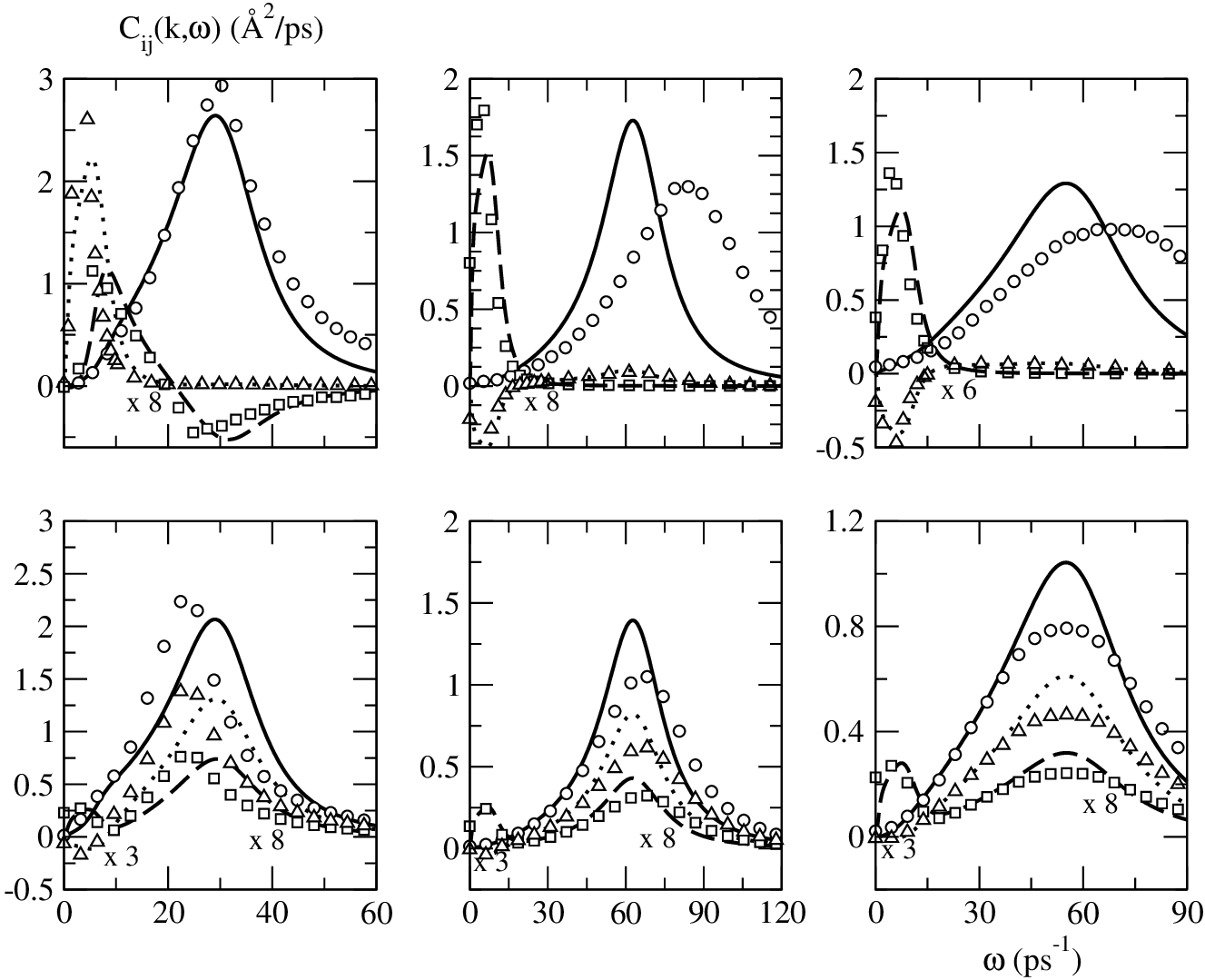}
\caption{Longitudinal current correlation functions for Li$_4$Pb. 
Lines are theoretical results and symbols are simulation data. 
Upper pannel: Ashcroft-Langreth partials. Lower pannel: 
Bhatia-Thornton partials. 
The meaning of the lines and symbols is the same as in the previous figure.
The $k$ values for the different columns are the same as in fig.~\ref{fijkt_LiPb}.
\label{ckw_LiPb}} 
\end{figure}

Finally we address the longitudinal current correlation functions. In fig.~\ref{ckw_LiMg} we show the AL and the BT functions for \LiMg, and in fig.~\ref{ckw_LiPb} those corresponding to Li$_4$Pb. Again the case of \LiMg is very similar to pseudobinary Li.
For small $k$, the three AL partials have a peak at a common frequency. 
On the other hand, 
$C_{12}(k,\omega)$ exhibits both a maximum and a minimum for all $k$, 
whose positions coincide with the maxima (or shoulders at small $k$)
of $C_{11}(k,\omega)$ and $C_{22}(k,\omega)$ respectively.
The behavior of the BT currents is also similar to pseudobinary Li: $C_{NN}(k,\omega)$ only shows one peak, and $C_{cc}(k,\omega)$ also shows only one peak, although since the decoupling number-concentration is not complete, a clear feature (a shoulder) does also appear at the frequency of the maximum of $C_{NN}(k,\omega)$.
In the case of Li$_4$Pb it is evident that the smallest $k$ 
shown in the graph is already outside 
the hydrodynamic region. For all the $k$ values shown we observe 
two characteristic frequencies, a small one for which we have a maximum in $C_{22}(k,\omega)$ and a maximum (at low $k$) or a minimum (at larger $k$) for $C_{12}(k,\omega)$, and a high one at which there is a maximum in $C_{11}(k,\omega)$ and a minimum (at small $k$) or a maximum (at larger $k$) in $C_{12}(k,\omega)$.
The BT currents, on the other hand, always show a common maximum at the high characteristic frequency and a maximum in $C_{cc}(k,\omega)$, minimum in $C_{Nc}(k,\omega)$, at the low one.
We therefore see that in Li$_4$Pb $C_{cc}(k,\omega)$ has a clear double peak structure, each situated close to the low and high frequency roots of the model.

\section{Conclusions.}

We have presented an extension of Lovesey's viscoelastic model to binary mixtures that correctly recovers
the same model for the one-component system in the case of pseudobinary mixtures, including both the
collective and the single-particle properties. 

We have observed in the MD simulations of this kind of system a clear appearance of propagating modes in the single-particle dynamic properties of the one-component system, specifically in the longitudinal self-current correlation function, which the viscoelastic model reproduces correctly. 

In the case of alloys, the appearance of two propagating roots in the model 
leads naturally to the fast sound phenomenon, which is explained as the dominance of the high frequency propagating contribution in the DSF of the light component, appearing near the $k$ values where the amplitudes of the two propagating roots coincide. For smaller $k$ the amplitude of the high frequency component goes to zero and therefore all the DSFs have a component with a peak at a common frequency, namely that of the smaller propagating root. Whether this component shows up as a peak or not in the DSFs, depends strongly on the amplitude (and width) of the diffusive contributions, which in some cases (like \LiMg) can cover the propagating contribution.

The viscoelastic model is able to reproduce the different behaviors of the ISFs of pseudobinary Li and \LiMg on one hand and Li$_4$Pb on the other. The extremely slow decay of the ISFs for the former two cases is dictated by the small value and large amplitude associated to the smallest real root of the model, which moreover in the case of pseudobinary Li is strictly the single-particle contribution. In the case of Li$_4$Pb a combination of factors leads to a more rapid decay: first, the smallest real root is larger than in \LiMg ; second, the amplitudes associated to this root in the 22 and 12 partials are much smaller than those in \LiMg, due to the behavior of the amplitudes which follow the corresponding partial structure factors; and third, in the case of the 11 partial it is the larger real root the dominant one for small $k$, leading therefore to a much faster overall decay.
Another important difference is the presence in Li$_4$Pb of propagating concentration modes, which are absent in \LiMg. The presence (or absence) of clear side peaks in $S_{cc}(k,\omega)$ is neatly reproduced by the viscoelastic model,
in spite of its very small magnitudes.

We conclude by stressing that although quantitative differences do appear between the results of the model and the simulations, nevertheless the model reproduces accurately the main characteristic features of the dynamic properties of the systems studied, including very fine details as the peaks of $S_{cc}(k,\omega)$ and even unexpected peaks in the single-particle properties. 
We believe that this warrants a wide application of the model to the analysis of data (possibly obtained by any other means) on the dynamic properties of alloys, and on the single-particle properties of one-component systems, which will complement the already usual application of the model for the collective dynamics of one-component systems.
 
\begin{acknowledgments}
We acknowledge finnancial support from DGICYT (MAT2002-04393-C0201 and 
BFM2003-08211-C0302), Junta de Castilla y Leon (VA073-02), 
the NSF (Career Award 0133504) and the Army Research Office.
\end{acknowledgments}

\appendix*
\section{}

Here, we derive the expressions for the matrices ${\bf \Xi}$ and 
${\bf K}(\kk)$ in the case of a two-component system.  
In Section \ref{theory}, these matrices were determined by the integral  

\begin{displaymath}
\label{Ap1} 
\int_0^{\infty} dt \; \exp \left[ - \frac{t^2}{2} 
{\bf N}(\kk, t=0) \right]
\end{displaymath}

\noindent  The exponential is easily performed after diagonalizing ${\bf N}(\kk, t=0)$. Denoting  by \{$\lambda_1(\kk)$, $\lambda_2(\kk)$\} and 
\{${\bf x}_1(\kk), {\bf x}_2(\kk)$\}         
its eigenvalues and eigenvectors we can write 

\begin{equation}
{\bf N}(\kk, t=0)  = 
{\bf P}(\kk)\; {\bf N_D}(\kk)\; {{\bf P}(\kk)}^{-1} 
\end{equation}

\noindent where ${\bf N_D}(\kk)$ is a diagonal matrix having the eigenvalues 
\{$\lambda_1(k)$, $\lambda_2(k)$\} on the diagonal, whereas ${\bf P}(\kk)$ 
is a matrix having 
\{${\bf x}_1(k), {\bf x}_2(k)$\}  as its column vectors.
The exponential is then straightforward and the wanted integral is readily evaluated 
leading to  

\begin{equation}
\label{Ap3}
\tilde {\bf M}(\kk, z=0) \; = \; \frac{\sqrt \pi}{2} \; {\bf \Xi} \; 
{\bf P}(\kk) \; {\bf D}(\kk) \; {{\bf P}(\kk)}^{-1} \; {\bf M}(\kk, t=0) 
\end{equation}

\noindent where ${\bf D}(\kk)$ is a diagonal matrix with elements 
\{$\sqrt { 2/ \lambda_1(k)}$, $\sqrt {2/ \lambda_2(k)}$\} 
on the diagonal. Now, by using eqn. (\ref{Fkz}), it is obtained that 

\begin{eqnarray}
& & \tilde {\bf F}(\kk, z=0) = \nonumber \\
& & \frac{2}{\sqrt \pi} \left[ {\bf \Xi} 
{\bf P}(\kk) {\bf D}(\kk) {\bf P}(\kk)^{-1}
{\bf M}(\kk, t=0) \right]^{-1} {\bf F}(\kk, t=0) \nonumber \\
\end{eqnarray}
 
\noindent and by examining the $k \to \infty$ limit of this equation,  
the elements of the matrix ${\bf \Xi}$ are obtained.  
In particular we have 

\begin{eqnarray}
\lambda_i(\kk \to \infty) & = & \frac{2 k^2 k_B T}{m_i} \nonumber    \\
{\bf P}(\kk \to \infty) & = & {\bf I} \nonumber 
\\
\left[{\bf M}(\kk \to \infty, t=0)\right]_{ij} & = & \delta_{ij} \frac{k^2 k_B T}{m_i} \nonumber \\
\left[\tilde{\bf F}(\kk \to \infty, z=0)\right]_{ij} & = & \pi S_{ij}(\kk, \omega=0) = \nonumber \\
& = & \delta_{ij} \left( \frac{\pi}{2  k^2} \frac{m_i}{k_B T} \right)^{1/2}
\end{eqnarray}

\noindent leading to ${\bf \Xi} = (2 \sqrt {2}/ \pi) {\bf I}$. 
 
Now, we turn to the determination of the matrix 
${\bf K}(\kk)$.  By inserting eqn. (\ref{Ap3}) in eqn. (\ref{Krelax}), 
and using the previous result for ${\bf \Xi}$,  
it is obtained that 

\begin{equation}
{\bf K}(\kk) =  
\; \frac{\sqrt \pi}{2} \;  
{\bf P}(\kk) \; {\bf D}(\kk) \; {{\bf P}(\kk)}^{-1}  
\end{equation}
 
\noindent The explicit expressions for the matrix elements of 
${\bf K}(\kk)$ are

\begin{eqnarray}
K_{11}(\kk) & = & \frac{2}{\sqrt \pi} 
\frac{N_{12}(\kk)N_{21}(\kk) \lambda_1^{1/2}(\kk) + 
\beta^2(\kk) \lambda_2^{1/2}(\kk)}
{N_{12}(\kk)N_{21}(\kk) + \beta^2(\kk)} \nonumber \\
K_{12}(\kk) & = & \frac{2}{\sqrt \pi} 
\frac{N_{12}(\kk) \beta (\kk) [ \lambda_1^{1/2}(\kk) -  
 \lambda_2^{1/2}(\kk)] }
{N_{12}(\kk)N_{21}(\kk) + \beta^2(\kk)} \nonumber \\
\nonumber \\
K_{21}(\kk) & = & \frac{2}{\sqrt \pi} 
\frac{N_{21}(\kk) \beta (\kk) [ \lambda_1^{1/2}(\kk) -  
 \lambda_2^{1/2}(\kk)] }
{N_{12}(\kk)N_{21}(\kk) + \beta^2(\kk)} \nonumber \\
K_{22}(\kk) & = & \frac{2}{\sqrt \pi} 
\frac{N_{12}(\kk)N_{21}(\kk) \lambda_2^{1/2}(\kk) + 
\beta^2(\kk) \lambda_1^{1/2}(\kk)}
{N_{12}(\kk)N_{21}(\kk) + \beta^2(\kk)}
\nonumber \\
\end{eqnarray}

\noindent where the $N_{ij}(\kk)$ are the matrix elements of 
${\bf N}(\kk, t=0)$, the eigenvalues are given by

\begin{eqnarray}
\lambda_1(\kk) & = & \frac{1}{2} [ N_{11}(\kk)+ N_{22}(\kk) ] \nonumber \\
 & + & \left[ \left( \frac{ N_{11}(\kk) - N_{22}(\kk)}{2} \right) + 
N_{12}(\kk) N_{21}(\kk) \right]^{1/2} \nonumber \\
\nonumber \\
\lambda_2(\kk) & = & \frac{1}{2} [ N_{11}(\kk)+ N_{22}(\kk) ] \nonumber \\
 & - &  \left[ \left( \frac{ N_{11}(\kk) - N_{22}(\kk)}{2} \right) + 
N_{12}(\kk) N_{21}(\kk) \right]^{1/2} \nonumber \\
\end{eqnarray}

\noindent and $\beta (\kk) = \lambda_1(\kk) - N_{11}(\kk)$.

\clearpage

\end{document}